%% file: ms.tex
\documentclass{emulateapj}
\usepackage{epsfig}

\input defs.tex

\input epsf
\bibpunct{(}{)}{;}{a}{}{,}

\begin{document}

\lefthead{BINARY QUASARS AT HIGH REDSHIFT} \righthead{HENNAWI \etal}
\title{Binary Quasars at High Redshift I: 24 New Quasar Pairs at ${\rm
    z} \sim 3-4$}

\author{Joseph F. Hennawi\altaffilmark{1,2,3,4},
  Adam D. Myers\altaffilmark{4,5},
  Yue Shen\altaffilmark{6},
  Michael A. Strauss\altaffilmark{6},
  S. G. Djorgovski\altaffilmark{7},
  Xiaohui Fan\altaffilmark{8},
  Eilat Glikman\altaffilmark{7},
  Ashish Mahabal\altaffilmark{7},
  Crystal L. Martin\altaffilmark{9},
  Gordon T. Richards\altaffilmark{10},  
  Donald P. Schneider\altaffilmark{11},
  Francesco Shankar\altaffilmark{12}
}

\altaffiltext{1}{NSF Astronomy and Astrophysics Postdoctoral Fellow}
\altaffiltext{2}{Department of Astronomy, University of California at
  Berkeley, 601 Campbell Hall, Berkeley, CA 94720-3411}
\altaffiltext{3}{Max-Planck-Institut f\"ur Astronomie, K\"onigstuhl
  17, D-69117 Heidelberg, Germany}
\altaffiltext{4}{Visiting Astronomer, Kitt Peak National Observatory,
  National Optical Astronomy Observatory, which is operated by the
  Association of Universities for Research in Astronomy (AURA) under
  cooperative agreement with the National Science Foundation.}
\altaffiltext{5}{Department of Astronomy, University of Illinois at 
  Urbana-Champaign, Urbana, IL 61801}
\altaffiltext{6}{Princeton University Observatory, Peyton Hall,
  Princeton, NJ 08544}
\altaffiltext{7}{Astronomy Department, California Institute of Technology, Pasadena, CA 91125}
\altaffiltext{8}{Steward Observatory, University of Arizona, 933 North Cherry Avenue, Tucson, AZ 85721}
\altaffiltext{9}{Department of Physics, University of California, Santa Barbara, CA 93106}
\altaffiltext{10}{Departmentof Physics, Drexel University, 3141 Chestnut 
  Street, Philadelphia, PA19104 }
\altaffiltext{11}{Department of Astronomy and Astrophysics, Pennsylvania State
University, 525 Davey Laboratory, University Park, PA 16802}
\altaffiltext{12}{Max-Planck-Institüt f\"ur Astrophysik,
  Karl-Schwarzschild-Str. 1, D- 85748, Garching, Germany.}

\begin{abstract}
  The clustering of quasars on small scales yields fundamental
  constraints on models of quasar evolution and the buildup of
  supermassive black holes. This paper describes the first systematic
  survey to discover high redshift binary quasars. Using
  color-selection and photometric redshift techniques, we searched
  $8142$~deg$^2$ of SDSS imaging data for binary quasar candidates,
  and confirmed them with follow-up spectroscopy. Our sample of 27
  high redshift binaries (24 of them new discoveries) at redshifts
  $2.9 < z < 4.3$ with proper transverse separations $ 10~{\rm kpc} <
  R_{\perp} < 650~{\rm kpc}$ increases the number of such objects
  known by an order of magnitude.  Eight members of this sample are
  very close pairs with $R_{\perp} < 100~{\rm kpc}$, and of these
  close systems four are at ${\rm z} > 3.5$. The completeness and
  efficiency of our well-defined selection algorithm are quantified
  using simulated photometry and we find that our sample is $\sim
  50\%$ complete. Our companion paper uses this knowledge to make the
  first measurement of the small-scale clustering ($R < 1~\hMpc$
  comoving) of high-redshift quasars.  High redshift binaries
  constitute exponentially rare coincidences of two extreme ($M
  \gtrsim 10^9~\msol$) supermassive black holes. At $z \sim 4$ there
  is about one close binary per $10~{\rm Gpc}^3$, thus these could be
  the highest sigma peaks, the analogs of superclusters, in the early
  Universe.
\end{abstract}


\section{Introduction}
\label{sec:intro}

Astronomers do not yet understand how quasars fit into the galaxy
formation hierarchy and how they relate to the underlying structure
and evolution of the dark matter.  In the current paradigm, every
massive galaxy is hosted by a dark matter halo and is thought to have
undergone a luminous quasar phase; in this picture quasars at high
redshift are the progenitors of the local dormant supermassive black
holes found in the centers of all nearby bulge-dominated
galaxies. This fundamental connection is supported by the tight
correlations observed between the masses of central black holes and
the velocity dispersions (or masses) of their old stellar populations
\citep{Magorrian98,FerrMerr00,Gebhardt00,Tremaine02,HaringRix04} and by the fact
that the black hole mass density in the local Universe is commensurate
with the luminosity density produced by quasars at high redshift
\citep[e.g.][]{Soltan82,SB92,yt02,HCO04,marconi04,Shankar04,Shankar08a,
  Shen-th09}.

Quasars are likely to reside in massive hosts \citep{Turner91} and it
has been suggested that they occupy the rarest peaks in the initial
Gaussian density fluctuation distribution
\citep{ER88,CK89,NS93,Djor99a,Djor99b}. It is also
thought that quasar activity is triggered by the frequent galaxy
merger which are a generic consequence of bottom up structure
formation
\citep{Carl90,HR93,BarnHern96,Bahcall97,Thacker06,Hopkins08,Wetzel09a,
Wetzel09b},
and merger-based models provide a good description of many of the
observed aggregate properties of the quasar population
\citep{HL98,CV00,kh00,WL02,Thacker06,Hopkins06,Thacker08,Shen-th09}.

In a hierarchical Universe, the large-scale clustering of a population
of objects is directly related to their host dark matter halo masses
\citep{CK89,MW96}. Hence measuring quasar clustering probes the
relationship between quasar luminosity and halo mass, providing
constraints on the quasar lifetime or duty cycle \citep[][but see
Wyithe \& Loeb 2008]{CK89,HH01,MW01}, physical parameters governing
black hole fueling such as the radiative efficiency and Eddington
ratio \citep{WyitheLoeb05,Shankar08b,Shen-th09}, and the scatter in these
parameters \citep{WMC08}.  The advent of large spectroscopic quasar
surveys has led to firm constraints on the clustering
\citep{pmn04,croom05,Myers06,Myers07a,daAngela08,Paddy08,Shen08,Ross09}
of quasars at ${\rm z}\lesssim 2.5$ -- like luminous galaxies, quasars show
little apparent clustering evolution with redshift. In addition, the
first estimates of the dependence of clustering on luminosity
\citep{PN06,Shen08}, redshift \citep{Shen07}, the presence of broad
absorption line (BAL) troughs \citep{ShenBAL08}, and radio brightness
\citep{Shen08} have been made. The cross-correlation of quasars with
galaxies has also resulted in complementary constraints
\citep{AS05a,AS05b,Serber06,Coil07,Paddy08}. 

The clustering of high redshift quasars is much more challenging to
measure \citep{Kundic97,Stephens97} because they are intrinsically
rare.  But the Sloan Digital Sky Survey \citep[SDSS;][]{York00} has
recently enabled a measurement of quasar clustering on large scales at
${\rm z}>3$. \citet{Shen07} measured the clustering of quasars from $2.9 \le
z \le 5.4$ using a sample of $\gtrsim 4000$ quasars selected from the
SDSS DR5 \citep{DR5}. Dividing their sample into two redshift bins,
they measured comoving correlation lengths ($\gamma = 2$) of $r_0 =
16.9\pm 1.7~\hMpc$ and $r_0 = 24.3\pm 2.4~\hMpc$, for $2.9 \le z \le
3.5$ and ${\rm z} > 3.5$, respectively.  Thus ${\rm z} > 3$ quasars cluster much
more strongly than their ${\rm z}\lesssim 2.5$ counterparts ($r_0 \simeq
7~\hMpc$) , implying that high-redshift quasars are highly biased and
inhabit very massive ($M\gtrsim 10^{13}~\hmsol$) dark matter halos.
These large scale clustering constraints at low and high redshift now
play a fundamental role in our understanding of the quasar evolution
and the buildup of supermassive black holes
\citep{Thacker06,Lidz06,HopkinsLidz07,Thacker08,Djor08,Hopkins08,WMC08,
  Shankar08b,Shen-th09}.

Quasar clustering on small scales ($R < 200~{\rm kpc}$ proper) yields
independent and complementary constraints on quasar evolution.
\citet{Djor91} first suggested that the handful of binary quasars
known at the time implied a factor of $\sim 100$ larger clustering
amplitude on scales $R < 50~{\rm kpc}$ than predicted by naively
extrapolating the quasar correlation function power law from larger
scales, and proposed that this was due to the enhancement of quasar
activity during merger events.  \citet{BINARY} conducted an extensive
follow-up campaign to find companions of SDSS quasars, providing the
first measurement of the quasar correlation function on scales
$10~\hkpc < R_{\rm prop}< 400~\hkpc$. For $R_{\rm prop} \lesssim
50~{\rm hkpc}$, they detected an order of magnitude higher clustering
than the expectation from the larger-scale quasar correlation function
extrapolated as a power law, providing compelling evidence that the
quasar correlation function steepens on sub-Mpc scales. This result
has since been confirmed with better accuracy using more homogeneous
selected photometric \citep{myers07b} and spectroscopic samples
\citep{myers08}, albeit with less excess over a power law. Additional
circumstantial evidence for enhanced small scale clustering comes from
the recent discovery of the first triple quasar
\citep{Djor07}. Similarly large small-scale clustering may be present
around low-redshift AGN at ${\rm z} < 0.6$ \citep[][but see Padmanabhan et
al. 2008]{Serber06,Strand08}. Motivated by these small scale
measurements, several theoretical investigations have argued that
excess small scale clustering occurs naturally if quasar activity is
triggered by mergers \citep{Thacker06,Hopkins08,Wetzel09a,Thacker08}
and have considered the constraints that small scale clustering
measurements place on how quasars populate their dark matter halos
\citep{Wetzel09a,Shankar09}.



Measuring small scale quasar clustering is of particular interest at
high redshifts (${\rm z} > 3$), where the population of supermassive black
holes powering quasars is rapidly growing and the large scale
clustering indicates that quasars are hosted by extremely massive dark
matter halos ($M\gtrsim 10^{13}~\hmsol$), which are the progenitors of
today's rich clusters.  Could the mechanism which triggers the growth
of these extreme black holes ($M\gtrsim 10^{9}~\msol$ at high
redshift) be different from that at lower redshifts? Are ${\rm z}\sim 4$
quasars always at the center of their host dark matter halo, or can
they be hosted by less massive satellites? If a high redshift binary
quasar represents an exponentially rare coincidence of two extremely
massive black holes (and dark matter halos/subhalos), are these the
highest sigma peaks, the analogs of superclusters, in the early
Universe?  The evolution of small scale clustering to high redshift
can shed light on all of these questions. However, even in the large
${\rm z} > 3$ quasar sample studied by \citet{Shen07}, the smallest scale at
which the correlation function can be measured is $\sim 2~\hMpc$. The
reason for this is twofold. First, close quasar pairs with angular
separations $\lesssim 60\arcsec$ ($\sim 1.5~\hMpc$ at ${\rm z}\sim 4$), are
extremely rare --- the mean quasar separation at ${\rm z} > 3$ is $\sim
150~\hMpc$ and at small separations, the correlation function does not
increase as fast as the volume decreases.  Second, because of the
finite size of the optical fibers of the SDSS multi-object
spectrograph, only one quasar in a close pair with separation
$<55\arcsec$ can be observed on a given plate\footnote{An exception to
  this rule exists for a fraction ($\sim 30\%$) of the area of the
  SDSS spectroscopic survey covered by overlapping plates.  Because
  the same area of sky was observed spectroscopically on more than one
  occasion, there is no fiber collision limitation for these regions.}
\citep{sdss-targ}. The only sub-arcminute high redshift quasar pair in
the literature is a $33\arcsec$ pair of quasars at ${\rm z}=4.25$ discovered
serendipitously by \citet{Schneider00}. \citet{Djor03} discovered a
quasar at ${\rm z}=5.02$ $196\arcsec$ away from the high redshift quasar at
${\rm z}=4.96$ discovered by \citet{Fan99}, corresponding to a proper
transverse separation of $900~\hkpc$. This is the highest redshift
pair of quasars known.





In this paper, we present the results of a systematic survey to
discover high redshift binary quasars and measure the small-scale
clustering of quasars at ${\rm z}\sim 3-4$ for the first time. Using
color-selection and photometric redshift techniques, we searched the
SDSS photometric data for binary quasar candidates, which we confirmed
spectroscopically with follow-up observations.  The outline of this
paper is as follows. In \S~\ref{sec:selection}, we present our
algorithm for finding high redshift binary quasars candidates in the
SDSS photometric data, and quantify the completeness and efficiency of
our search. Our spectroscopic follow-up observations are described in
\S~\ref{sec:observations}, and the binary quasar sample is presented
in \S~\ref{sec:sample}, where we also summarize the status of our
survey.  We summarize and conclude in \S~\ref{sec:conc}. In a
companion paper \citep{Shen09} (henceforth Paper II), we use the
binaries presented here to measure the small-scale ($R < 1~\hMpc$
comoving) clustering of high redshift quasars.

Throughout this paper we use the best fit \emph{Wilkinson Microwave
  Anisotropy Probe} cosmological model of \citet{wmap05}, with
$\Omega_m = 0.26$, $\Omega_\Lambda =0.74$, $h=0.7$.  Because both
proper and comoving distances are used in this paper, we will always
explicitly indicate proper distances (quoted in kpc). It is helpful to
remember that in the chosen cosmology, for a typical quasar redshift
of ${\rm z}=3.5$, an angular separation of $\Delta\theta=1\arcsec$
corresponds to a proper transverse separation of $R_{\perp}= 7.6~{\rm
  kpc}$ (comoving $R_{\perp}=24.1~\hkpc$), and a velocity difference
of $1000~\kms$ at this redshift corresponds to a proper radial
redshift space distance of $s_{\parallel}=2.9~{\rm Mpc}$ (comoving
$s_{\parallel}=9.1~\hMpc$). All magnitudes quoted are SDSS asinh
magnitudes \citep{Lupton99}. To avoid confusion between redshift and
$z$-band magnitude, we will always use a text ``z'' to denote redshift.

\section{Finding Binary Quasars}

\label{sec:selection}

Because binary quasars at ${\rm z} \gtrsim 3$ are extremely rare, we must
target candidates to $i=21$, as faint as the SDSS imaging data will
allow, to build up statistics.  This flux limit is significantly
fainter than the SDSS spectroscopic flux limit ($i < 20.2$), and
because fiber collisions exclude close pairs in SDSS spectroscopy, we
must select binary candidates from the imaging data and then follow
them up spectroscopically.  Finding binary quasars thus amounts to
constructing a \emph{photometric} quasar catalog with a high
completeness and efficiency at faint magnitudes.

Our algorithm for selecting ${\rm z} > 3$ binary quasars evolved somewhat
over the course of our survey. We experimented with different
approaches, such as searching for companions only around
spectroscopically confirmed quasars following \citet{BINARY},
searching for pairs in the photometric quasar catalog of
\citet{Richards08} \citep[see also][]{Richards04}, and constructing
our own photometric quasar catalog.  The final algorithm adopted is an
amalgam of these methods and we describe it in detail in what
follows. In \S~\ref{sec:SDSS}, we provide details of the SDSS
spectroscopic quasar sample.  Many of our binary candidates have one
member bright enough for the spectroscopic survey, and these data are
used whenever possible. The photometric quasar catalog of
\citet{Richards08} is described in \S~\ref{sec:KDE}. This catalog did
not perform well enough at high redshift to be our sole selection
algorithm, however we utilize it in an auxiliary manner to improve the
efficiency of our search.  An important step in understanding the
completeness of our binary selection was the construction of simulated
quasar data, as we describe in \S~\ref{sec:simulate}. The core of our
selection algorithm is the construction of a photometric quasar
catalog which carefully treats photometric errors, which we introduce in
\S~\ref{sec:locus} and apply to the SDSS imaging data in \S~\ref{sec:photo}.
The final algorithm for selecting binaries is
presented in \S~\ref{sec:binary}, where we also estimate the
completeness of our search -- a necessary step for the small scale
quasar clustering constraints presented in our companion paper
\citep{Shen09}.

\subsection{The SDSS Spectroscopic Quasar Sample}
\label{sec:SDSS}

The Sloan Digital Sky Survey uses a dedicated 2.5m telescope
\citep{Gunn06} and a large format CCD camera \citep{Gunn98} at the
Apache Point Observatory in New Mexico to obtain images in five broad
bands \citep[$u$, $g$, $r$, $i$ and $z$, centered at 3551, 4686, 6166,
7480 and 8932 \AA, respectively;][]{Fuku96,Stoughton02} of high
Galactic latitude sky in the Northern Galactic Cap.  The imaging data
are processed by the astrometric pipeline \citep{astrom} and
photometric pipeline \citep{photo}, and are photometrically calibrated
to a standard star network
\citep{Smith02,Tucker06,ubercal}. Additional details on the SDSS data
products can be found in the data release papers
\citep[e.g.][]{DR1,DR2,DR3}.

Quasar candidates are targeted for follow-up spectroscopy based on
colors measured from the SDSS imaging data \citep{qsoselect}. These
candidates along with other targets (i.e. galaxies, stars,
serendipity), are observed with two double spectrographs producing
spectra covering \hbox{3800--9200 \AA} with a spectral resolution
ranging from 1800 to 2100.  Details of the spectroscopic observations
can be found in \citet{York00}, \citet{Castander01}, and
\citet{Stoughton02}. Quasars observed through the Fifth Data Release
\citep{DR5} have been catalogued by \citet[][see also Schneider et
al. 2005]{qsocat-DR5}.

High-redshift quasars are optically unresolved and thus cannot be
distinguished from stars based on image morphology.  The majority of
quasar candidates are selected based on their location in the
multidimensional SDSS color-space. All magnitudes are reddening
corrected following the prescription in \citet{SFD98}. Objects with
colors that place them far from the stellar locus and which do not
inhabit specific `exclusion' regions (e.g., places dominated by white
dwarfs, A stars, and M star-white dwarf pairs \citep{Smolcic04}) are
identified as primary quasar candidates.  An $i$ magnitude limit
of~19.1 is imposed for candidates whose colors indicate a probable
redshift of less than~$\approx$~3, while higher redshift candidates are
targeted if $i < 20.2$.  The colors of quasars become difficult to
distinguish from F-stars at ${\rm z} \approx 2.7$
\citep{Fan99-colors,Richards01}, which significantly lowers the
efficiency of quasar target selection. The SDSS quasar target selection
therefore deliberately sparse samples in that region of color space,
resulting in high incompleteness \citep{qsoselect,Richards06} near
this redshift.  We therefore focus our high redshift binary quasar
search at ${\rm z} \ge 2.9$.


\subsection{The SDSS KDE Photometric Sample}
\label{sec:KDE}

\citet{Richards04} and \citet{Richards08} have constructed faint
($i\lesssim 21$) photometric samples of quasars from the SDSS
photometry alone, by separating quasars from stars using knowledge of
their relative densities in color space. A technique known as Kernel
Density Estimation (henceforth KDE, or `the KDE technique') uses
training sets of quasars and stars to obtain non-parametric estimates
of their respective densities in color space.  Then a probability can
be assigned for any photometric object of interest for being
quasar-like, or star-like. Objects with quasar-like probability above
a threshold are selected as photometric quasars, and are assigned a
photometric redshift, and a probability that the photometric redshift
is correct \citep{Weinstein04}.  The most recent \citet{Richards08}
photometric quasar catalog of (henceforth the DR6
KDE catalog) covers the 8417 ${\rm deg}^2$ area corresponding to the
SDSS Data Release 6 \citep[DR6;][]{DR6}, extends down to an extinction
corrected flux limit of $i<21.3$, and attempts to classify quasars
over the full range of redshifts accessible to the SDSS imaging, $0 <
z \lesssim 5.5$.

The fundamental drawback of the KDE approach is that the probabilities
of an object being quasar-like or star-like do not take into account
photometric errors. At faint magnitudes where errors become large, the
KDE catalog suffers from incompleteness which is not currently
understood. Since a quantitative estimate of the completeness of our
selection is necessary to place constraints on small-scale quasar
clustering, using the KDE catalog alone as our selection algorithm is
insufficient. Of the 27 ${\rm z} > 2.9$ binary quasars in the
SDSS footprint (see Table~\ref{table:sample}), 25 have both members
bright enough to be in the KDE catalog. Of these, 17 have both
members in the KDE catalog and the other eight have only one member in the
KDE.


\subsection{Simulated Binary Quasars}
\label{sec:simulate}


Our binary survey extends to $i = 21$, and because of photometric
errors, the completeness decreases to fainter magnitudes.  Hence we
need to generated simulated binary quasars to determine our
completeness, as we do not have a large sample of spectroscopically
confirmed ${\rm z} > 2.9$ quasars at magnitudes fainter than the SDSS quasar
survey flux limit of $i=20.2$.


It will prove to be convenient if the simulated quasars have an
approximately correct redshift and number count distribution. Even at
bright magnitudes, the redshift and magnitude distribution of the SDSS
quasar sample has the selection function of the targeting algorithm
imprinted on it.  In particular, the SDSS spectroscopic sample suffers
high incompleteness at ${\rm z} = 2.7-3.0$ and at ${\rm z} \simeq 3.4$ \citep[see
Figure~8 of][]{Richards06}.  Instead, we determine the redshift and
number count distribution from the measured quasar luminosity function
(LF). Usable luminosity functions for quasars at redshift $2.9<z<4.5$
which probe the faint apparent magnitudes ($i<21$) include
\citet{Wolf03}, \citet{Jiang06}, \citet{Richards06}, and
\citet{HopkinsGTR}. After comparing all of these results, we decided
on a combination of the \citet{Jiang06} LF fits (at ${\rm z}<3.5$) and the
COMBO-17 PDE \citep{Wolf03} fits (at ${\rm z}>3.5$) for the model LF, and
scaled the result to our standard cosmology. We refer the reader to
our companion paper \citep{Shen09} for additional details of the
adopted LF.

Our procedure for simulating binary quasars is then as follows:

\begin{itemize}
\item[--] Draw a quasar redshift ${\rm z}$ from the redshift distribution
  $dN\slash d{\rm z}(i<21)$ determined by integrating the LF. 
\item[--] Given this redshift, draw two magnitudes, $i_1$ and $i_2$,
  for the two members of the binary from the apparent magnitude
  distribution of quasars  at this redshift, $n({\rm z},< i)$.
\item[--] If $i_1 \le 20.2$, locate a quasar in the real SDSS quasar
  sample with a similar value of $i$ and ${\rm z}$. Use the real quasar
  photometry for the first quasar.
\item[--] If $i_1 > 20.2$, then locate a quasar in the real SDSS
  quasar sample with well measured colors ($i < 19.5$) and similar
  ${\rm z}$. Make the real quasar fainter by scaling all five 
  fluxes $f^m$, $m = (u,g,r,i,z)$ , by the implied ratio of $i$-band
  fluxes.  Add Gaussian noise to the scaled fake data with standard
  deviation $\sigma^m(f^m)$ consistent with the SDSS noise model (see below). 
\item[--] Use same procedure to generate photometry for the second member
of the binary. 
\end{itemize}

Note that because the aforementioned procedure is based on real quasar 
photometry at a given redshift, the effect of Ly$\alpha$ forest and Lyman-limit
system absorption on quasar colors are automatically modeled correctly. 

To model the noise in the photometry as a function of flux level
in filter $m$, $\sigma^m(f^m)$, we randomly select $10^6$ stellar
objects from the SDSS photometric data and apply a running median
filter. Given a five-dimensional ($ugriz$) flux vector $f^m$ for a
simulated quasar, we then add random Gaussian noise.  Note that our
procedure assumes no covariance between the flux noise in different
filters, which is a good approximation for point sources
\citep{Scranton05}.  As an example, a randomly chosen simulated quasar
at ${\rm z}=3.2$ with $i=20.2$ has $m=(23.00,20.66,20.32,20.20,20.00)$
and $\sigma = (0.47,0.025,0.026,0.034,0.082)$, where these values
represent real photometric data (since $i\le 20.2$). Similarly at
${\rm z}=4$ we simulate an object with $m =
(24.68,21.58,20.28,20.20,20.21)$ and $(0.87,0.058,0.027,0.033,0.12)$.
For quasars at the same redshifts but with $i=21$, again chosen at
random, we have example objects with
$m=(23.27,21.42,20.91,21.00,20.83)$ and $\sigma
=(0.66,0.053,0.047,0.072,0.21)$ at ${\rm z}=3.2$, and
$(25.92,22.83,21.12,21.00,21.10)$ and $\sigma =
(0.70,0.17,0.055,0.072,0.26)$ at ${\rm z}=4$.

\subsection{Locus Distances}
\label{sec:locus}

\subsubsection{The Quasar Locus}
\label{sec:qsolocus}
Although quasars have a wide range of luminosities, the majority of
unobscured quasars have similar optical/ultraviolet spectral energy
distributions.  \citet{Richards01} demonstrated that most quasars
follow a tight color-redshift relation in the SDSS filter system, a
property which has been exploited to calculate photometric redshifts
of quasars \citep{photoz,Budavari01,Weinstein04}. It is thus possible
to efficiently target binary quasars by searching for pairs of objects
with similar, quasar-like colors \citep{thesis,BINARY}.  But stars are
a significant contaminant and must be avoided: for $i < 19$ stars
outnumber quasars on the sky by a factor of $\gtrsim 50$
\citep[e.g.]{Yasuda01,Sesar07} even at high galactic latitudes,
although stars are less dominant at fainter magnitudes (factor $\sim 25$
for $i < 21$).  The vast majority of stars detected by SDSS are on the
main sequence \citep[$>98\%$;][]{Finlator00,Helmi03,Juric08} and form
an extremely tight well defined temperature sequence \citep{Ivezic04},
referred to as the stellar locus in color-color diagrams.  Hence
quasars can be selected efficiently by targeting unresolved objects
near the quasar locus but far from the stellar locus, except at
redshift ${\rm z} \approx 2.7$ and ${\rm z} = 3.4$ where the two loci cross
\citep{Fan99-colors,Richards01,Richards06}.

As objects become progressively fainter, photometric measurement
errors blur the distinction between the quasar and stellar loci. The
SDSS quasar spectroscopic target selection algorithm \citep{qsoselect}
is not well-suited for such faint objects ($i\sim 21$), and as
discussed in \S~\ref{sec:KDE} the DR6 KDE catalog is not well
characterized at these faint magnitudes either.  Hence, we introduce a
new algorithm for selecting high redshift quasars focusing on a proper
statistical treatment of photometric errors. Following \citet{BINARY},
we define a statistic that quantifies the likelihood that an
astronomical object has colors consistent with a given model, based on
the mean quasar color-redshift relation and the stellar locus.

Given the fluxes of an object in the five SDSS bands $f^m_{\rm data}$,
we ask whether they are consistent with proportionality to a
model
\be 
f^m_{\rm data} = A f^m_{\rm model}, 
\ee 
where $f^m_{\rm model}$ is a five-dimensional vector of model fluxes, and this
relationship holds for a single proportionality constant in all bands. For the 
quasar locus, the model will be the average run of scaled quasar flux 
with redshift $f^m_{\rm model}({\rm z})$, whereas for the stellar locus the 
model  will be the scaled stellar flux as a function of surface 
temperature. 

For a quasar, the maximum likelihood value of the parameters $A$ and
${\rm z}$, given measured fluxes $f_{\rm data}^m$, can be determined
by $\chi^2$ minimization: 
\be \chi^2(A,{\rm z}) = \sum_{\rm
  ugriz}\frac{[f^m_{\rm data}- A f^m_{\rm model}({\rm z})]^2}
{[\sigma^m_{\rm data}]^2 + A^2 [\sigma^m_{\rm model}({\rm z})]^2},
\label{eqn:chi}
\ee 
where $\sigma^m_{\rm data}$ are the photometric measurement
errors, and $\sigma^m_{\rm model}({\rm z})$ is the intrinsic $1\sigma$
scatter about the mean quasar flux-redshift relation, both presumed to
be Gaussian distributed. The assumption of Gaussian statistics is
valid for the measurement errors \citep{Scranton05}, but
\citet{Richards01} found a red tail in the distribution of quasar
colors about the mean, which is due to dust extinction
\citep{Richards03,Hopkins04}.



To compute $f^m_{\rm model}({\rm z})$ we follow \citet{Richards01} to
fit the mean color-redshift relation for quasars, but instead of
fitting colors we rescale all the $f^m$ to have the same value of $f^r
+ f^i + f^z$ (and $\sigma^m$ rescaled accordingly), and fit $f^m_{\rm
  model}({\rm z})$. This procedure is preferable to fitting colors for
two reasons. First, for high-redshift quasars that have dropped out of
a filter, say $u$-band, forming the $u-g$ color degrades a high
signal-to-noise ratio $g$-band measurement with the noisy $u$ band
measurement. Second, at low signal-to-noise ratios Pogson magnitudes
or SDSS asinh magnitudes \citep{Lupton99} pose several disadvantages,
such as bias in the resulting colors and non-Gaussian error
distributions, whereas the rescaled fluxes are immune from these problems. 

To fit $f^m_{\rm model}(z)$, we start with a sample of $\sim 77,000$
quasars in the SDSS DR5 quasar catalog \citep{qsocat-DR5}. Restricting
attention to quasars in the redshift range $0.3 \le z \le 5.5$,
removing broad-absorption line quasars (which would bias the mean
colors), and restricting attention to quasars with well measured
redshifts (see Shen et al. 2008) and the most accurate photometry,
results in a sample of 52,000 quasars, including 5143 quasars with ${\rm z}
> 2.9$ and 829 with ${\rm z} > 4$.  Fluxes are galactic extinction-corrected
following the prescription in \citet{SFD98} and rescaled to have the
same value of $f^r + f^i + f^z$. The rescaled flux in each band is
then fit as a function of redshift ${\rm z}$ using a 4th order
B-spline with a breakpoint spacing of $0.1$ and outlier rejection.
Given this mean flux-redshift relation $f^m_{\rm model}({\rm z})$, we
then determine the 1$\sigma$ scatter about it, $\sigma^m_{\rm
  model}({\rm z})$, by applying a running median filter to the
difference between the data and the fits, $f^m - f^m_{\rm model}(z)$,
sorted by redshift. By construction this median curve brackets $50\%$
of the data points and varies smoothly with redshift. The 1$\sigma$
scatter is then determined from the median curve by rescaling the
median by the appropriate factor to enclose 68\% of the data points
for a Gaussian distribution.  Our estimate of $\sigma^m_{\rm
  model}(z)$ is clearly approximate since it assumes Gaussian
statistics, and includes a significant contribution from photometric
errors from the fainter objects.

If the redshift of the quasar is known, minimizing $\chi^2$ with
respect to $A$ results in four degrees of freedom (the number of
independent colors that can be formed from five fluxes) and $\chi^2$
should follow the $\chi_4^2$ distribution. If instead ${\rm z}$ is
unknown, it can be estimated by minimizing $\chi^2$. However, the
resulting distribution of $\chi^2$ values will be close to $\chi_3^2$,
but not exactly because the colors are not a linear function of ${\rm
  z}$ \footnote{Recall that minimizing $\chi^2$ with respect to a
  parameter reduces the number of degrees of freedom by one only if
  $\chi^2$ is linear in that parameter.}.

\begin{figure*}[!t]
  \centerline{
    \epsfig{file=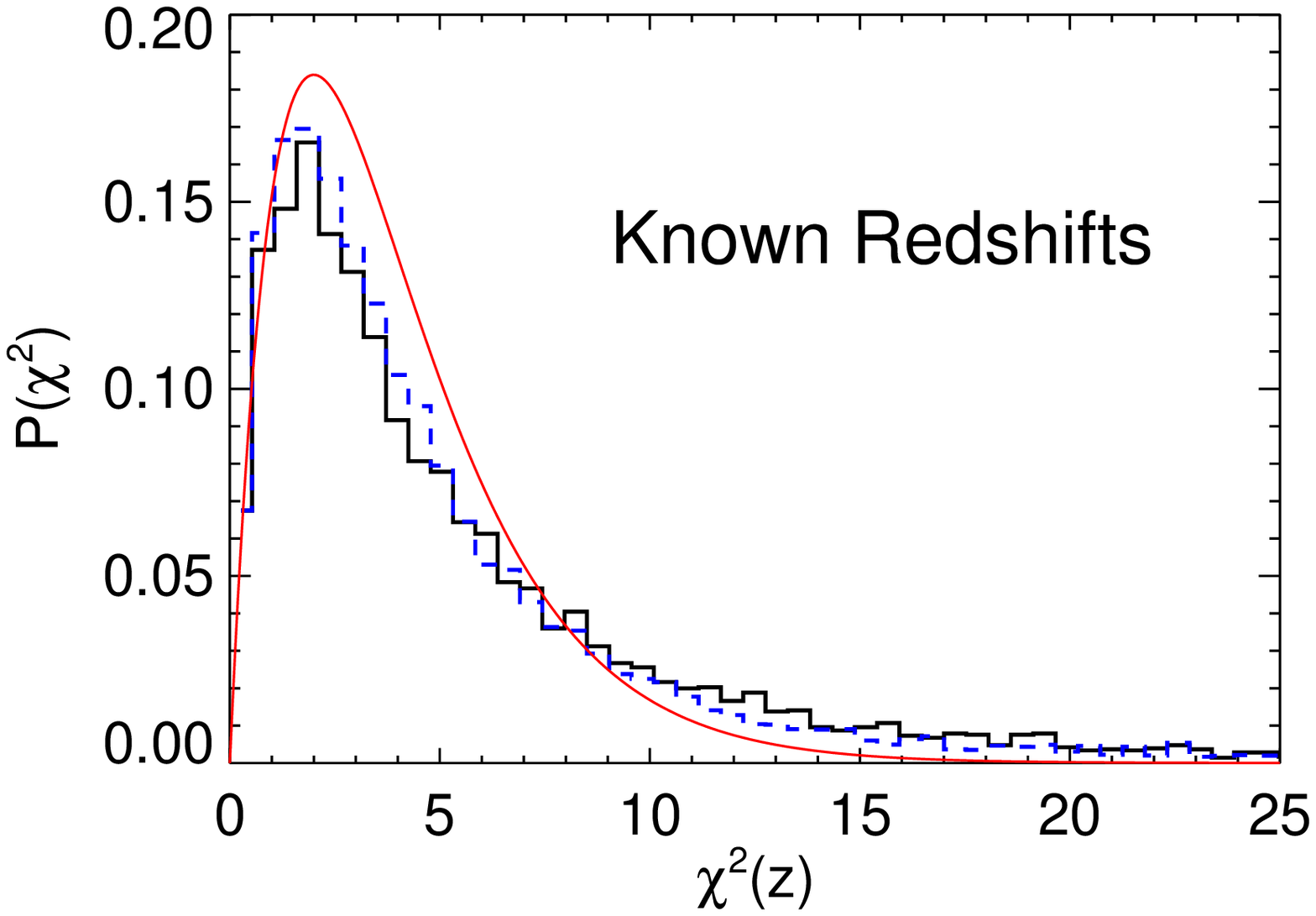,bb=0 0 504 360,width=0.50\textwidth}
    \epsfig{file=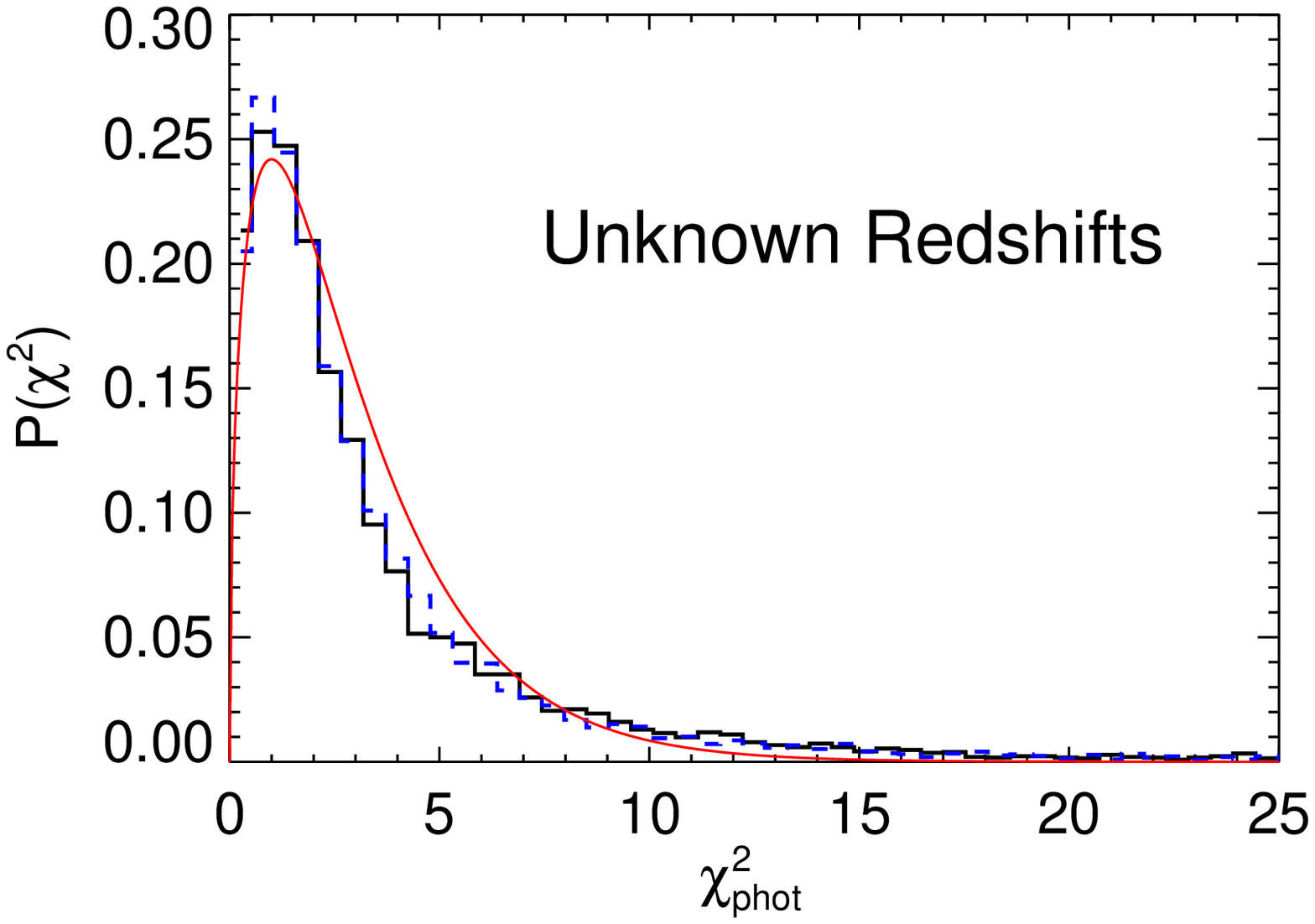,bb=0 0 504 360,width=0.50\textwidth}}
  \caption{ Histograms of quasar locus distance statistics. The left
    panel shows the distribution of $\chi^2({\rm z})$, where redshift
    was taken to be known and eqn.~\ref{eqn:chi} was minimized for $A$
    only.  The right panel shows the distribution of $\chi_{\rm
      phot}^2$, i.e. minimized with respect to both amplitude $A$ and
    redshift ${\rm z}$.  The solid (black) histograms are
    distributions for the 6696 $i < 20.2$ training sample quasars at
    ${\rm z} > 2.5$ used to fit the flux-redshift relation. The dashed
    (blue) histograms are from $\sim 600,000$ fainter ($i < 21$)
    simulated quasars at ${\rm z} > 2.5$.  The solid (red) curve shows the
    $\chi_4^2$ distribution in the left panel and the $\chi_3^2$
    in the left panel.\label{fig:X2_dist}}
\end{figure*}

The histograms in the left panel of Figure~\ref{fig:X2_dist} show the
resulting distribution of $\chi^2(z)$ when the $f^m_{\rm model}({\rm
  z})$ is evaluated at the true redshift of the quasar. The solid
(black) histogram is the distribution of $\chi^2$ for the subset of
the training sample used to fit the quasar flux-redshift relation,
which have $i < 20.2$ and ${\rm z} > 2.5$ (6696 quasars). The dashed (blue)
histogram is the distribution of $\chi^2$ for a sample of $\sim
600,000$ fainter simulated quasars with $i < 21$ (see
\S~\ref{sec:simulate}).  The solid (red) curve shows the $\chi_4^2$
distribution. The median value of $\chi^2$ for the real data is $3.93$
and it is $3.63$ for the simulated data, whereas the median of the
$\chi_4^2$ distribution is $3.37$. Our training set (solid histogram)
is broader than the $\chi_4^2$ distribution (solid curve) and exhibits
a tail toward large values, presumably because of the non-Gaussian tail 
of the quasar color distribution \citep{Richards01}.  For the fainter
simulated sample, the scatter about the mean becomes more dominated by
photometric measurement errors rather than intrinsic color
variations, and the distribution is closer to $\chi_4^2$. 

\begin{figure}[!t]
  \centerline{\epsfig{file=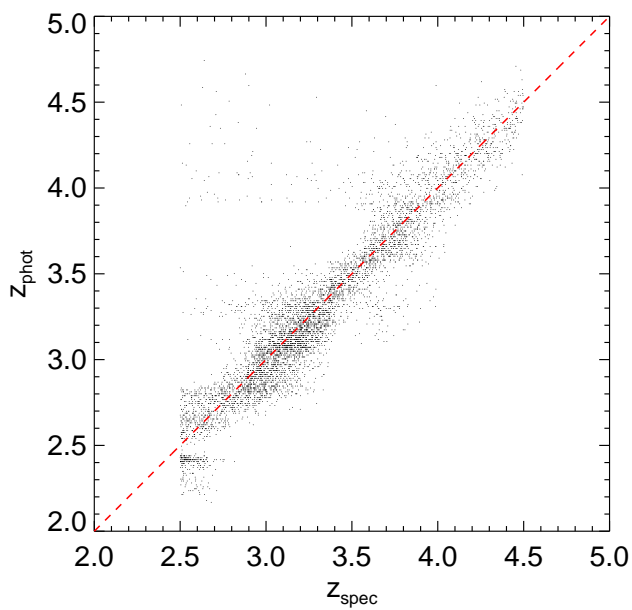,bb=0 0 504 504,width=0.5\textwidth}}
  \caption{Photometric redshift versus spectroscopic redshift for
    quasars with $2.5 < {\rm z} < 4.5$ in our training sample ($i
    \lesssim 20.2$). For ${\rm z} > 2.9$, $83\%$ of the training set
    quasars and $81\%$ of the fainter ($i< 21$) simulated quasars have
    $|{\rm z} - {\rm z}_{\rm phot}| < 0.2$.  The standard deviation
    (with outlier rejection) is $\sigma_{\rm z} = 0.14$ for the
    training set and $\sigma_{\rm z} = 0.15$ for the fainter simulated
    data. \label{fig:zphot}}
\end{figure}

The distribution of $\chi^2$ minimized with respect to both
parameters, $A$ and ${\rm z}$, is shown in the right panel of
Figure~\ref{fig:X2_dist}. This number quantifies the likelihood that a
quasar has photometric redshift ${\rm z}$, and we denote
it as $\chi_{\rm phot}^2$.  As expected, the resulting distributions
of $\chi_{\rm phot}^2$ are significantly narrower than before, with median 
values of $2.19$ and $2.17$ for the training sample and fainter simulations, 
respectively, which can be compared to the median of of the $\chi_3^2$ 
distribution, $2.38$.  The performance of our photometric redshift
algorithm for high redshift quasars is illustrated in
Figure~\ref{fig:zphot}, where we plot photometric redshift versus
spectroscopic redshift for quasars in the training sample with $2.5 < z < 
4.5$.  Our photometric redshifts are very accurate for high redshift
quasars, except and ${\rm z} < 2.7$ where we tend to
underestimate the true redshift. This degeneracy with lower
photometric redshifts is likely due to poor signal-to-noise ratio in the
$u$-band. The primary lever arm for ${\rm z} > 2$ photometric redshifts
comes from filters bracketing the Ly$\alpha$ forest spectral break,
which falls in $u$ for ${\rm z} < 2.7$.  For the quasars with $2.9 < z < 4.5$ of
interest for the binary selection in this paper, we find that $83\%$
of the real training set quasars and $81\%$ of the fainter simulated
quasars have $|z - z_{\rm phot}| < 0.2$.  Computing
the standard deviation with outlier rejection, we find that
$\sigma_{z} = 0.14$ for the training set data and $\sigma_{z} = 0.15$
for the fainter simulated data.


\subsubsection{The Stellar Locus}
\label{sec:starlocus}

We use a procedure similar to that used for the quasar locus to define
the stellar locus. First we isolate a set of representative stars with
accurate photometry using a set of SDSS spectroscopic plates on which
all point sources were targeted above a flux limit of $i<19.1$
regardless of color \citep[see][]{DR4}.  After applying cuts to remove
objects with noisy photometry and stars at the extremes of the color
distribution, we arrive at a set of 13,837 spectroscopically confirmed
stars with precise SDSS photometry.  We parametrize the path these
stars follow through the five-dimensional SDSS filter space using the
$g-i$ color, a decent proxy for stellar temperature.  Our fit
for $f^m_{\rm model}(g-i)$ uses the same methodology as we did for the
quasar locus: we rescale all stars to have the same value of
$f^r + f^i + f^z$, and each band $f^m$ is fit with a B-spline as a
function of $g-i$. The error $\sigma^m_{\rm model}(g-i)$ is more
subtle than for the stellar locus. The intrinsic width of the stellar
locus is so small \citep[$\sim 0.02$~mag][]{Ivezic04} that the scatter
in our star data about the mean relation is completely dominated by
photometric measurement errors, even at $i < 19.1$. Hence we simply
use $\sigma^m_{\rm model}(g-i)$ as a floor on the error to prevent
extreme values of $\chi^2$ for rare cases of extremely high
signal-to-noise ratio photometry.

\begin{figure*}[!t]
  \centerline{
    \epsfig{file=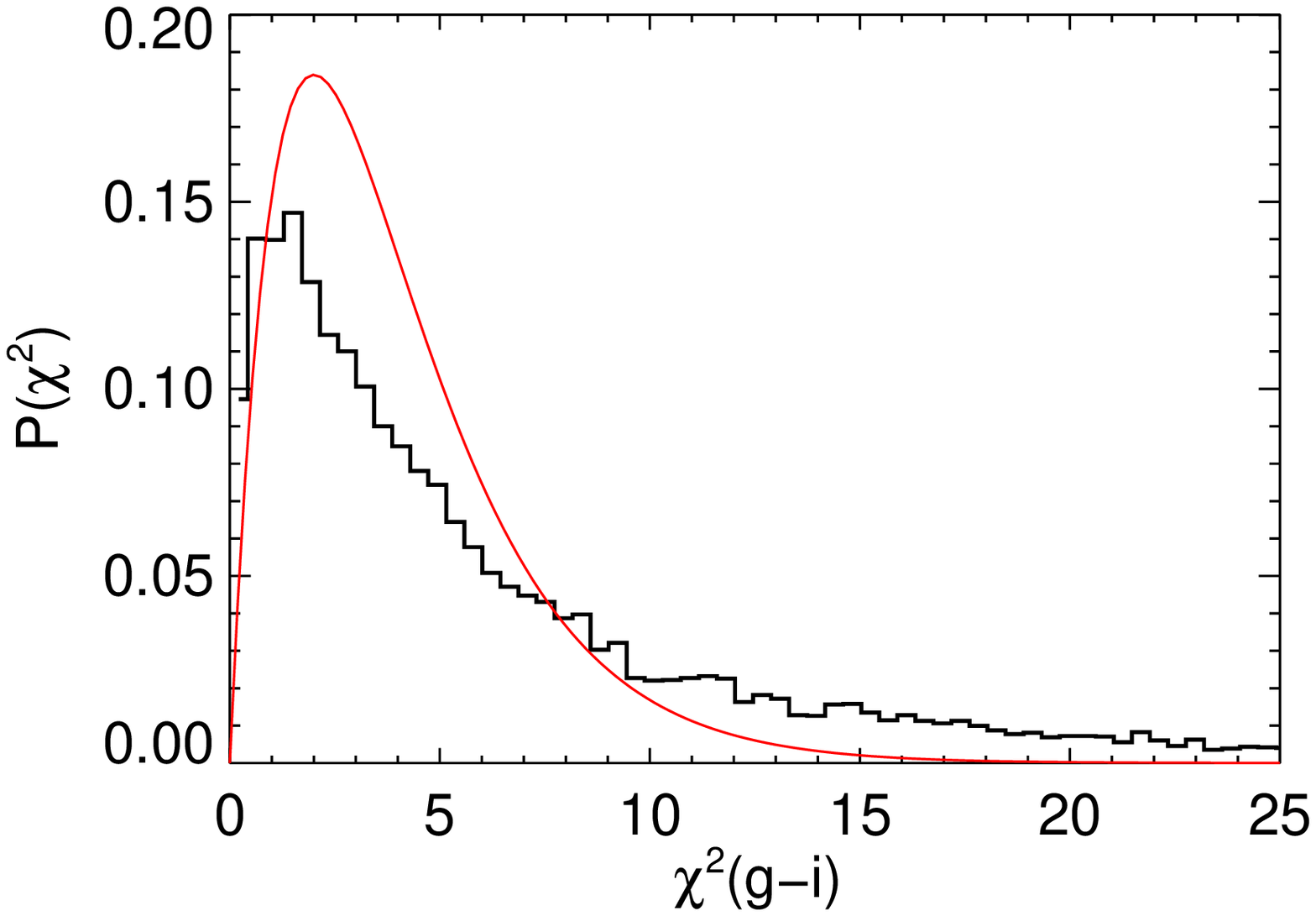,bb=0 0 504 360,width=0.50\textwidth}
    \epsfig{file=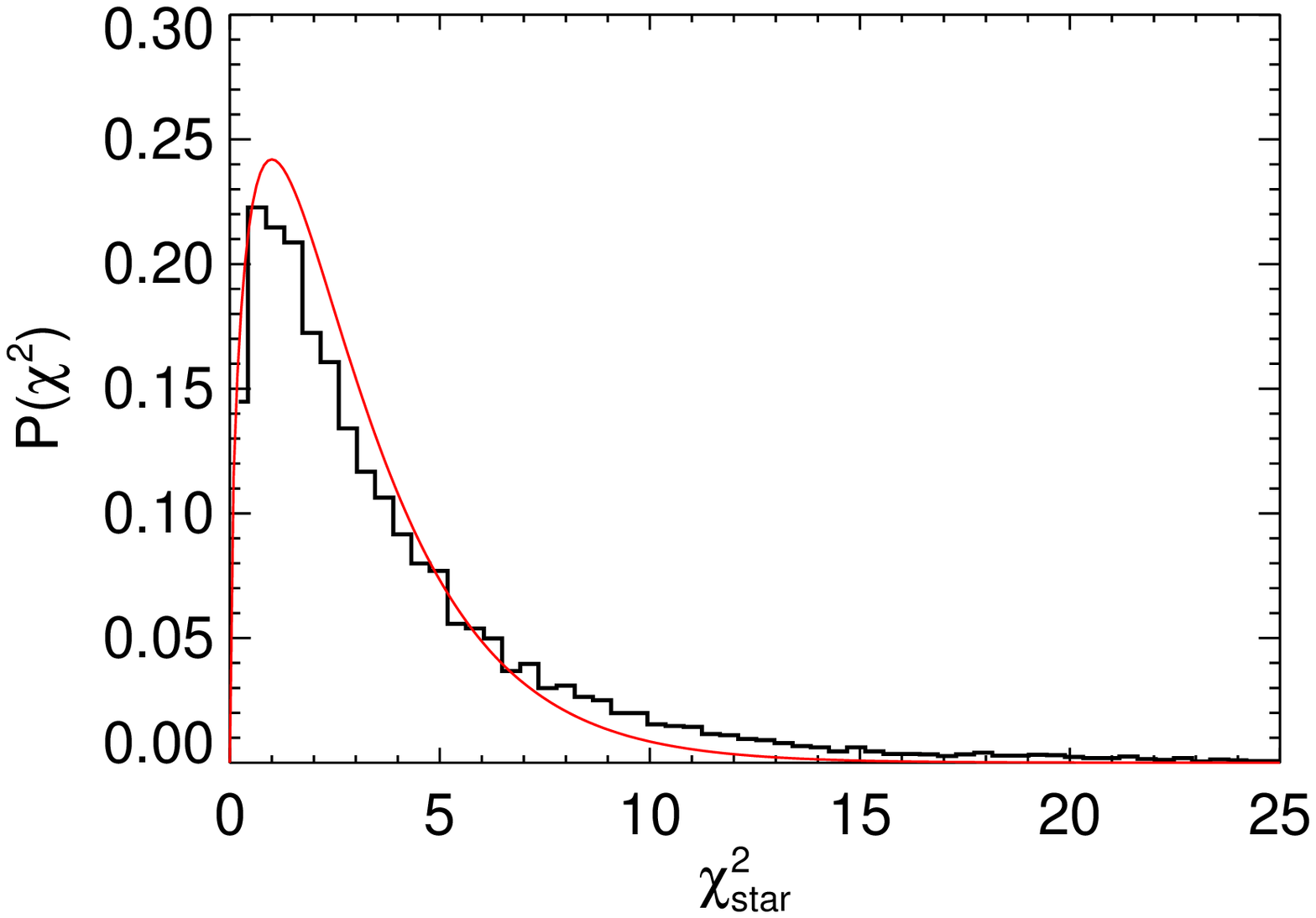,bb=0 0 504 360,width=0.50\textwidth}}
  \caption{ Histograms of stellar locus distance statistics for 13,837
    spectroscopically confirmed stars with $i<19.1$. The left panel
    shows the distribution of $\chi^2({g-i})$ if eqn.~\ref{eqn:chi} is
    minimized for amplitude $A$ only.  The right panel shows the
    distribution of the \emph{minimum distance} to the stellar locus,
    whereby $\chi^2(A,g-i)$ is minimized with respect to both
    parameters. The (red) curves shows the $\chi_4^2$ and $\chi_3^2$
    distributions in the left and right panels,
    respectively.\label{fig:X2_star}}
\end{figure*}

The histogram in the left panel of Figure~\ref{fig:X2_star} shows the
distribution of $\chi^2$ from the sample of stars which was used to
fit the stellar locus, where $f^m_{\rm model}(g-i)$ was evaluated at
the observed value of the $g-i$ for each star. The (red) curve shows
the $\chi_4^2$ distribution. The distribution of the data differs
significantly from $\chi_4^2$: the median value of the $\chi^2$ for
the training set is $4.35$ compared to a median of $3.37$ for the
$\chi_4^2$ distribution. This difference is due to a tail of stars
with atypical colors, probably arising from unusual metallicities or
surface gravities. In addition, intrinsic stellar colors depend on 
apparent magnitude, as the properties of stars change with distance, whereas
our fits ignore any such dependence. 

The \emph{minimum distance} to the stellar locus, which we designate
$\chi^2_{\rm star}$, can be computed by minimizing $\chi^2(A,g-i)$
with respect to both parameters. The statistical significance of an
outlier from the stellar locus is quantified by $\chi^2_{\rm
  star}$. The right panel of Figure~\ref{fig:X2_star} shows the
distribution of $\chi^2_{\rm star}$ compared to the $\chi_3^2$
distribution. Requiring $\chi^2_{\rm star}$ be larger than a 
specified value is an effective means of reducing stellar contamination
from quasar selection. 

\begin{figure*}[!t]
  \centerline{
    \epsfig{file=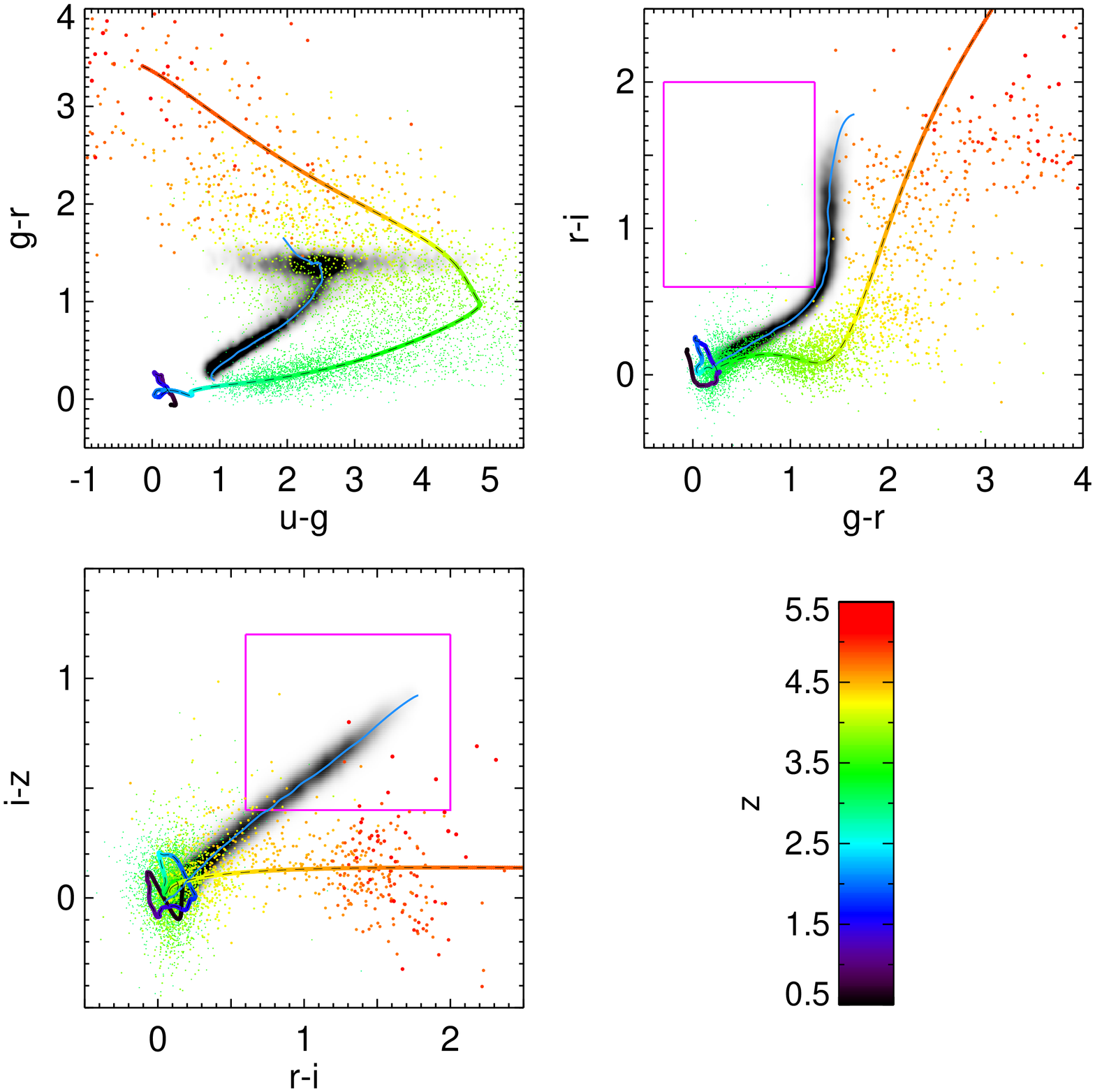,bb=20 0 675 690,width=1.0\textwidth}}
  \caption{Color-color diagrams illustrating the stellar locus and the
    quasar locus. The gray shaded region denotes stellar density for
    the $\sim 14,000$ stars used to fit the stellar locus, with darker
    regions corresponding to higher density.  The light blue curve is
    our fit to the stellar locus. The $\sim 5000$ ${\rm z} \ge 2.9$
    ($i\lesssim 20.2$) quasars in our training set are plotted as
    redshift color-coded points according to the color-bar at the
    lower right (lower redshift quasars are omitted for clarity).
    Higher redshift objects are plotted as larger points. The quasar
    locus is shown by the dashed black line, similarly color-coded to
    indicate redshift.  The magenta square box corresponds to a region
    excluded from SDSS quasar selection, which we similarly exclude
    (see \S~\ref{sec:photo}), because of contamination from white
    dwarf $+$ M-star pairs \citep[see][]{Richards01} \label{fig:col_plot}}
\end{figure*}

To visualize how quasars and stars objects move through the
four-dimensional SDSS color space ($u-g$, $g-r$,$r-i$,$i-z$), we plot
our fits to the quasar and stellar loci in the three color-color
diagrams in Figure~\ref{fig:col_plot}.

\subsection{Faint Photometric Quasar Selection}
\label{sec:photo}

\begin{figure*}[!t]
  \centerline{ \epsfig{file=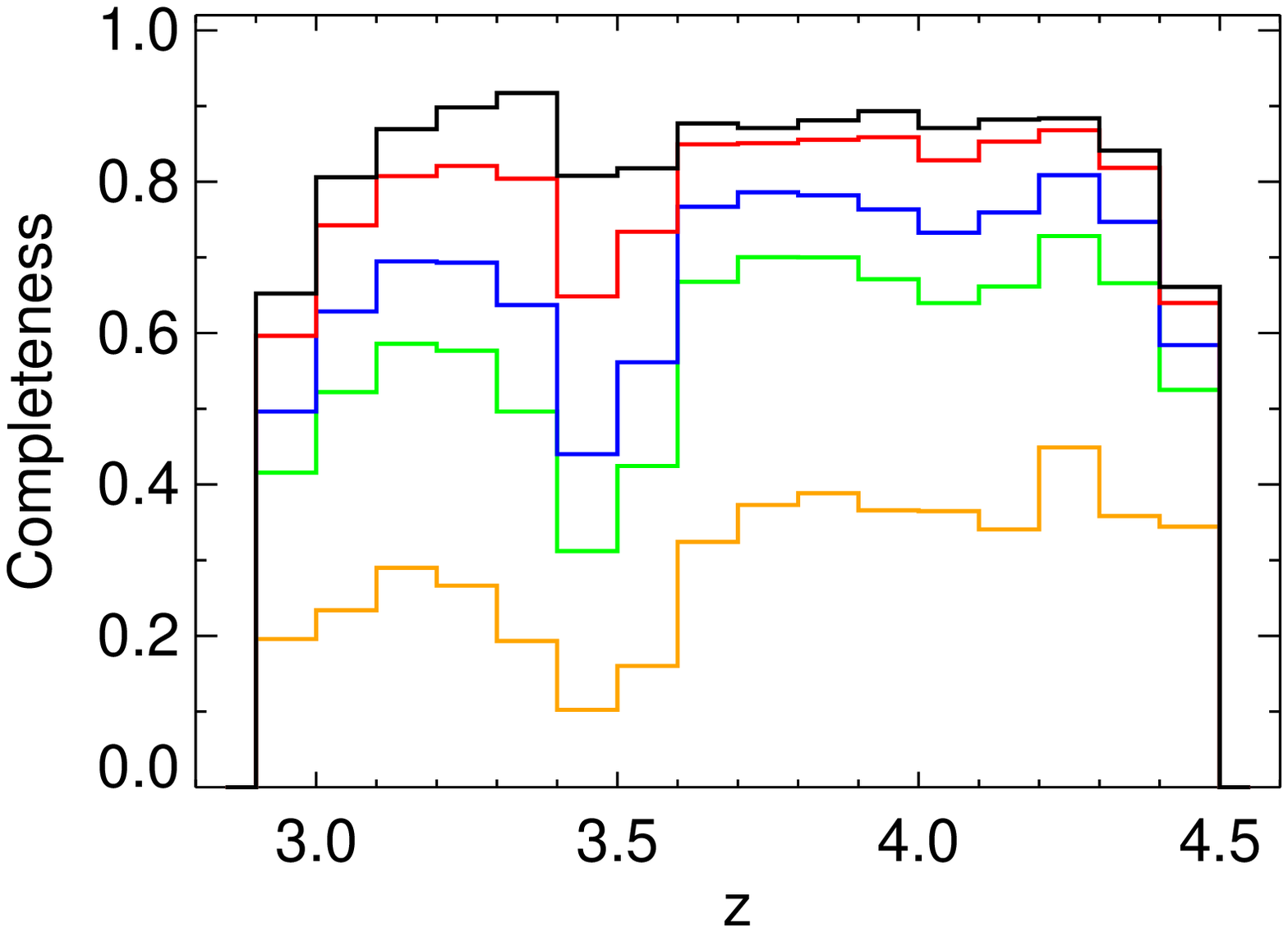,bb=0 0 504
      360,width=0.50\textwidth} \epsfig{file=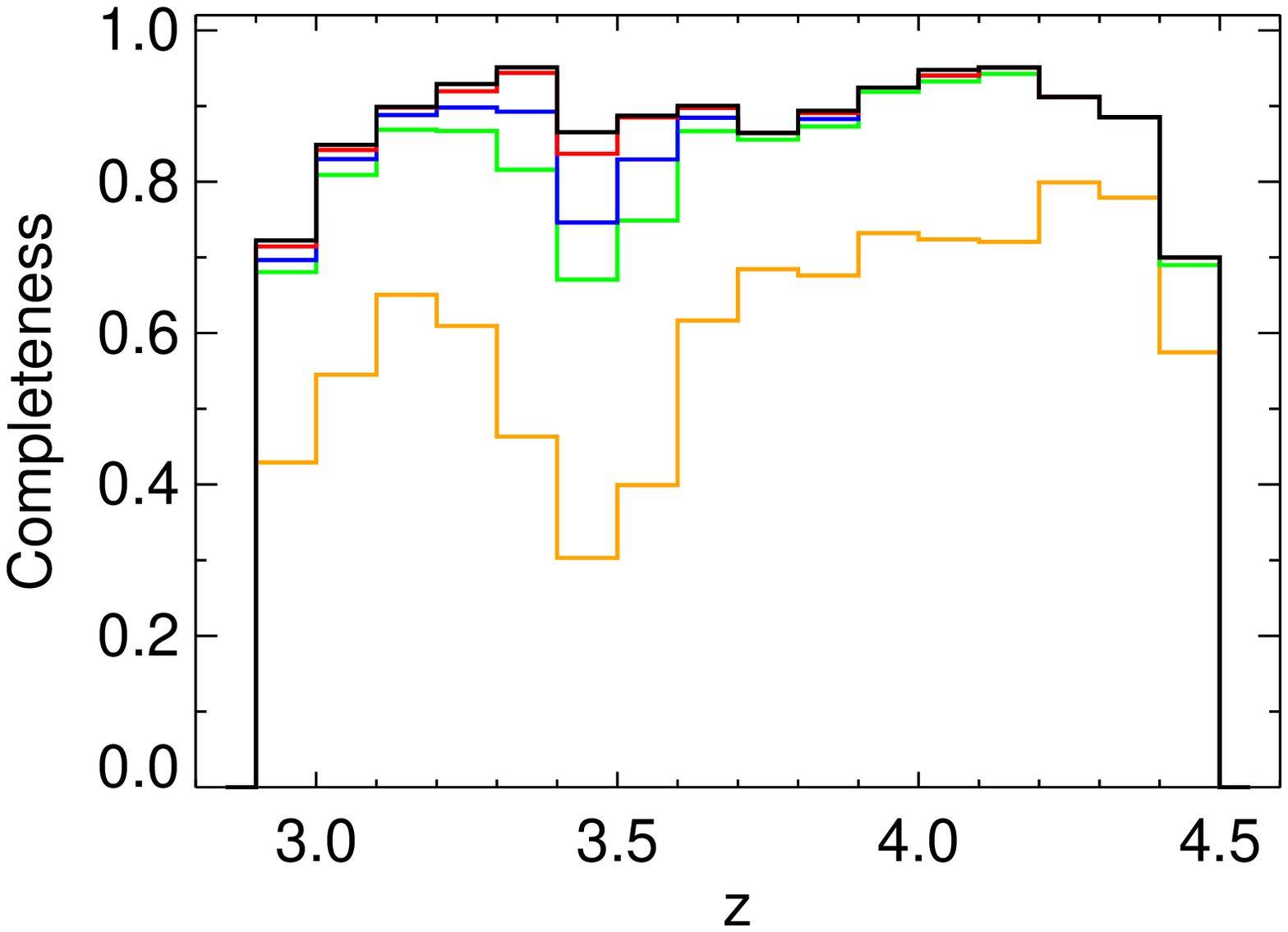,bb=0 0
      504 360,width=0.50\textwidth}} 
  \caption{Completeness of the faint photometric quasar catalog as a
    function of redshift for varying limits on $\chi^2_{\rm
      star}$. All criteria in \S~\ref{sec:photo} have been
    applied except criterion K on $\chi^2_{\rm star}$, which varies
    from top to bottom as $\chi^2_{\rm star} \ge [0,5,10,15,40]$.  The
    left panel shows completeness determined from $\sim 600,000$
    simulated quasars with $i < 21$. The right panel shows the
    completeness if we restrict to quasars with
    $i<20.2$. \label{fig:complete}}
\end{figure*}

We construct a photometric quasar catalog targeting objects consistent
with the quasar locus, that is small $\chi^2_{\rm phot}$, and
inconsistent with the stellar locus, hence large $\chi^2_{\rm
  star}$. Of our simulated $i < 21$ quasars with $2.9 < z < 4.5$,
$94\%$ have $\chi^2_{\rm phot} < 10$ (see Figure~\ref{fig:X2_dist}),
hence we can achieve a sample with high completeness by imposing this
cut on quasar locus distance.  To determine the placement of the cut
on stellar locus distance consider the left panel of
Figure~\ref{fig:complete}. The histograms show the completeness as a
function of redshift for varying limits on $\chi^2_{\rm star}$ as
determined from our $\sim 600,000$ simulated quasars with $i < 21$ and
${\rm z} \ge 2.9$.  Specifically, from top to bottom the five histograms
represent $\chi^2_{\rm star} \ge [0,5,10,15,40]$. In addition to the
cut on $\chi^2_{\rm star}$ we have imposed $\chi^2_{\rm phot} < 10$
and several other auxiliary cuts (described in detail below) which
have less of an impact on the overall completeness than $\chi^2_{\rm
  phot}$.

If there is no constraint on $\chi^2_{\rm star}$ (upper black curves
in Figure~\ref{fig:complete}) the other cuts result in a total
completeness of $82\%$ in the redshift interval $2.9 < z < 4.5$
Increasing the cut on $\chi^2_{\rm star}$ forces quasars to be higher
significance outliers from the stellar locus, thus decreasing the
completeness.  The decrease is particularly strong near ${\rm z}\simeq 3.4$,
because the stellar and quasar loci nearly overlap at this
redshift (see Figure~\ref{fig:col_plot}). Actually, the loci are still
distinct in the $ugr$ color-color diagram, but for fainter objects
photometric errors blur this distinction because of low
signal-to-noise ratio in the $u$-band. We choose $\chi^2_{\rm star} >
10$ (blue curves in Figure~\ref{fig:complete}) as a compromise between
completeness and the resulting efficiency of our search, which is
quantified below.

\begin{figure}[!t]
  \centerline{\epsfig{file=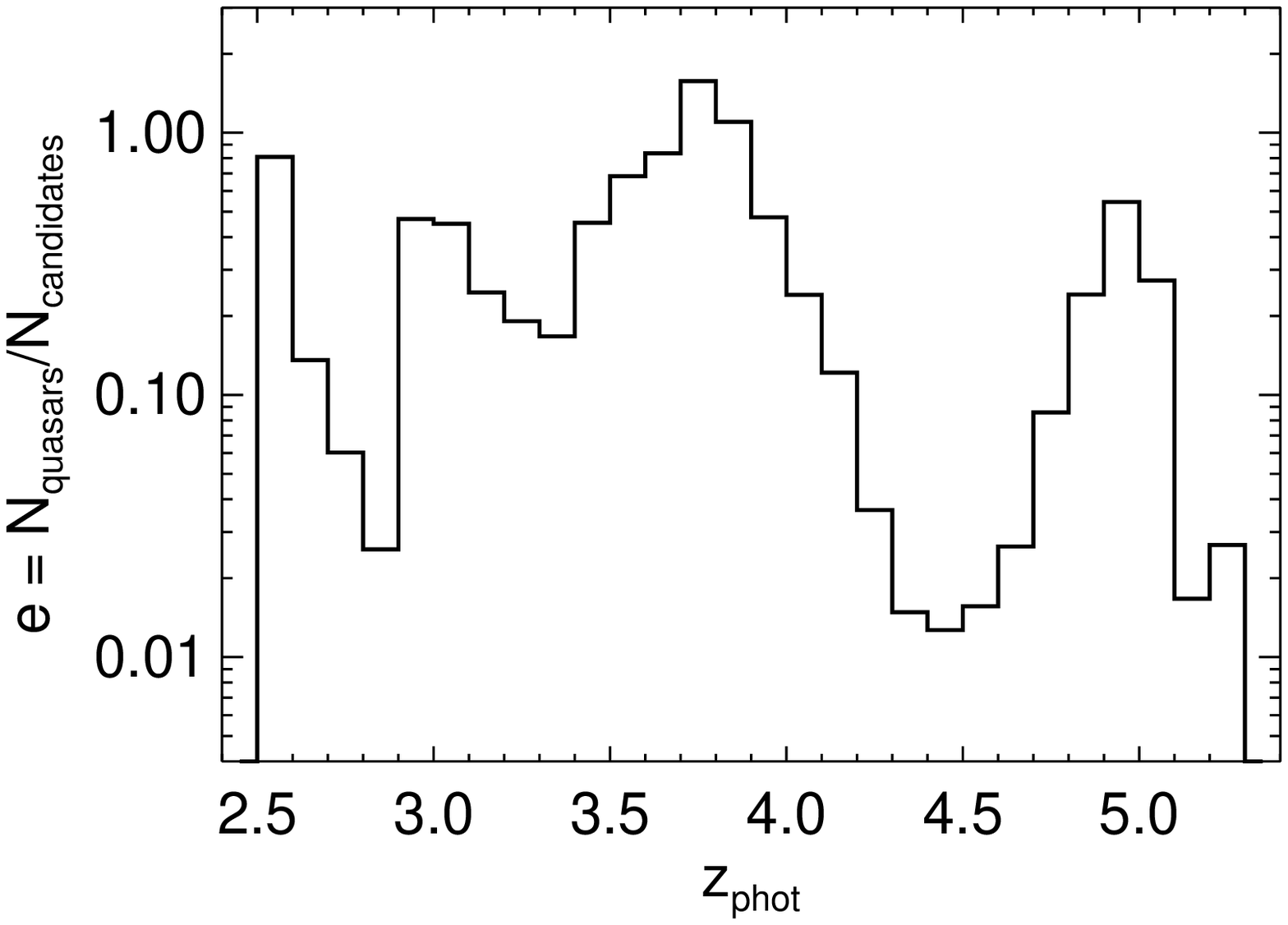,bb=0 0 504 360,width=0.50\textwidth}}
  \caption{Estimated efficiency of the faint photometric quasar catalog as
    a function of redshift. The criteria in S~\ref{sec:photo}
    have been applied, except that the photometric redshift range has been
    expanded to $2.5 < z_{\rm phot} < 5.3$.  For ${\rm z}\lesssim 4$,
    the selection efficiency is $\gtrsim 50\%$, except for dips at the
    problematic redshifts ${\rm z} \sim 2.8$ and ${\rm z}\sim 3.4$, where quasar 
    colors intersect the stellar locus (see Figure~\ref{fig:col_plot}). 
    The efficiency reaches 1.6, and is thus
    unphysically above unity at ${\rm z}_{\rm phot}\sim 3.8$, so we
    caution that our estimates are uncertain by at least this factor.
    For ${\rm z}_{\rm phot}\gtrsim 4.2$, the efficiency drops precipitously,
    reaching a minimum of $\sim 1\%$ at ${\rm z}\sim 4.4$ \label{fig:eff}}
\end{figure}


Our binary quasar selection algorithm contains several steps. The area
we consider for our search is the entire SDSS DR6 with an added
constraint on galactic latitude $|b| > 25^\circ$ to avoid
contamination from highly reddened sources, resulting in a final
survey area of $8142$~deg$^2$. In this area we first consider all
objects that fulfill the set of conditions below to be photometric
quasar candidates. In what follows we describe and motivate each
criterion and parenthetically quote the resulting completeness,
determined from our simulations ($i< 21$ and $2.9 < z < 4.5$), if each
criterion is applied sequentially:

\begin{itemize}

\item[A)] the image must be classified as STAR by the photometric
  pipeline ($>99\%$)

\item[B)] various cuts on flags produced by the SDSS
  photometric pipeline, similar to those used by spectroscopic quasar
  target selection \citep{qsoselect}, must be satisfied ($>99\%$)
  
\item[C)] The $i$-band photometric error must satisfy $\sigma_{\rm i} < 0.2$,
  excluding noisy data or spurious deblends ($>99\%$)

\item[D)] the galactic extinction in the $i$-band must be $A_{\rm i} <
  0.3$ reducing contamination from highly reddened objects ($>99\%$)

\item[E)] a photometric redshift dependent magnitude prior is imposed
  discarding stars which are stellar locus outliers but too bright to
  be quasars ($99\%$)

\item[F)] exclude objects with colors consistent with
  being white dwarf $+$ M-star pairs \citep[see
  Figure~\ref{fig:col_plot};][]{Richards01}, a significant contaminant
  ($99\%$)

\item[G)] the apparent magnitude must satisfy $i < 21$
  ($96\%$)\footnote{This reduction in completeness is due to Malmquist
    bias, given photometric errors.}

\item[H)] the $g-r$ color must be redward of the stellar locus in $g-r$
  vs $r-i$ color-color diagram (see Figure~\ref{fig:col_plot}) for
  $3.5 < z_{\rm phot} < 4.5$, significantly reducing contamination
  from faint red stars in this redshift range ($92\%$)
  
\item[I)] colors must be consistent with the quasar locus and have
  $\chi^2_{\rm phot} < 10$ ($88\%$)

\item[J)] the photometric redshift must be in the range $2.85 < {\rm
    z}_{\rm phot}$, thus excluding redshifts where selection is less
  efficient (see below, $82\%$)

\item[K)] colors must be inconsistent with the stellar locus and have
  $\chi^2_{\rm star} > 10$ ($64\%$)
\end{itemize}



Thus our final completeness for selecting individual quasars with $i <
21$ and $2.9 < z < 4.5$ is 64\%, with a redshift dependence given by
the blue line in the left hand panel of Figure~\ref{fig:complete}.  Our
selection is much more efficient for brighter higher signal-to-noise
ratio sources since stellar locus outliers can be identified at higher
statistical significance. The right panel of Figure~\ref{fig:complete}
shows completeness as a function of redshift for varying limits on
$\chi^2_{\rm star}$, but for brighter quasars with $i<20.2$ satisfying
the flux limit for the SDSS spectroscopic sample. The resulting
completeness for $i<20.2$ quasars is $83\%$, compared to 64\% for the
fainter $i<21$ flux limit. 





The efficiency of quasar targeting is another important criterion. Our
spectroscopic follow-up (see \S~\ref{sec:spectro}) does not allow us
to quantify this because we observed the best candidates first and our
follow-up is far from complete. In principle the efficiency could be
computed if we knew the number densities of the contaminants on the
sky. However, the number density of faint stars in the relevant
regions of color-space is not well constrained and is a strong
function of Galactic latitude.  Furthermore, faint red unresolved
galaxies may also be a significant contaminant and estimating their
number density is challenging.  Instead, we obtain a rough estimate of
the efficiency of our photometric quasar selection by comparing the
number density of quasars predicted from the luminosity function, to
the total number of sources identified in our catalog.

Specifically, we use the luminosity function (see discussion in
\S~\ref{sec:simulate}) to compute the number counts of quasars in
redshift bins, and multiply by the completeness determined from our
simulations (see Fig.~\ref{fig:complete}), to determine the total
number of quasars expected from our selection as a function of
redshift, $N_{\rm QSO}(z)$. To mimic the effect of photometric
redshift error we convolve $N_{\rm QSO}(z)$ with a Gaussian with
$\sigma_{z}=0.16$, the standard deviation of our photometric redshifts
for ${\rm z} > 2.5$ (see Figure~\ref{fig:zphot}). Then we can estimate the
efficiency $e \equiv N_{\rm QSO}(z_{\rm phot})\slash N_{\rm
  candidates}(z_{\rm phot})$, which provides a rough estimate of the
fraction of real quasars in our catalog as a function of photometric
redshift. 



The efficiency of our photometric catalog deduced in this way is shown
in Figure~\ref{fig:eff}. We have expanded the photometric redshift
range of our selection to $2.5 \le z_{\rm phot} \le 5.3$ for this
figure to illustrate the efficiency outside the limits of our survey
($2.9 < z < 4.5$), and to avoid edge effects in the photometric redshift
convolution. Otherwise the criteria applied are exactly those in
described in \S~\ref{sec:photo}.  For ${\rm z}\lesssim 4$, the photometric
selection efficiency is $\gtrsim 50\%$, except for dips at the
problematic redshifts ${\rm z} \sim 2.8$ and ${\rm z}\sim 3.4$. The efficiency
reaches 1.6, and is thus unphysically above unity at ${\rm z}_{\rm phot}\sim
3.8$, so we caution that our estimates are uncertain by at least this
factor.  For ${\rm z}_{\rm phot}\gtrsim 4.2$, the efficiency drops
precipitously, reaching a minimum of $\sim 1\%$ at ${\rm z}\sim 4.4$. The
contaminant at these redshifts is a rare population of red stars which
have colors distinct from the stellar locus defined in
\ref{sec:starlocus} (see Figure~\ref{fig:col_plot}), but which are
significantly more abundant than high redshift quasars.  Because these
red stars occur in much lower numbers than the stars on the stellar
locus, they were not included in our parametrization in
\S~\ref{sec:locus}. 

To improve the efficiency of our binary search we compare candidates
to the SDSS spectroscopic quasars, the FIRST radio survey
\citep{FIRST}, and the photometric KDE catalog \citep[see
\S~\ref{sec:KDE}][]{Richards08}, as we describe below.

\subsection{Binary Quasar Selection}
\label{sec:binary}


To select binary quasars we first apply the cuts described in 
\S~\ref{sec:photo} to the SDSS photometric data to identify an
initial photometric quasar catalog. We then search for pairs of
objects in this catalog that satisfy the following additional
criteria:

\begin{itemize}

\item[I)] the angular separation must be $\Delta < 120\arcsec$,
which projects to a comoving (proper) transverse distance of
$R_{\perp} = 2.7~\hMpc$ ($0.96~{\rm Mpc}$) and $R_{\perp} = 3.1~\hMpc$
($0.87~{\rm Mpc}$) at ${\rm z}=3$ and ${\rm z}=4$, respectively 

\item[II)] the quasar candidates must have similar photometric
  redshifts $\left|z_{\rm phot1} - z_{\rm phot2}\right| < 0.4$ (if
  applied to simulated quasar pairs ($2.9 < z < 4.5$) satisfying the
  criteria from \ref{sec:photo} only $4\%$ are rejected)

\item[III)] one member of the pair must either be: confirmed by the
  main SDSS spectroscopic survey to be a quasar at ${\rm z} > 2.9$, or
  a FIRST radio source \citep{FIRST}, or a member of the KDE
  photometric catalog \citep{Richards08}

\end{itemize}

In practice, the KDE completely dominates the auxiliary catalog
criterion III); the SDSS spectra and FIRST sources account for only
$\sim 2\%$ of matches. As we do not know the completeness of the KDE
catalog to $i \sim 21$, we cannot precisely quantify how III) impacts
our overall completeness for selecting binaries, but this uncertainty
has a very small effect, as we discuss in more detail below.



\begin{figure}[!t]
  \centerline{\epsfig{file=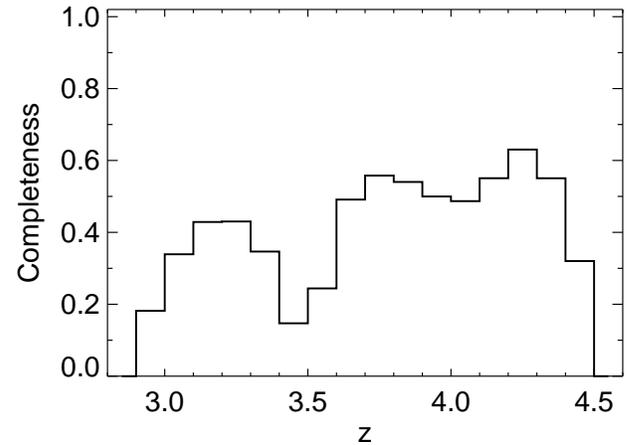,bb=0 0 504 360,width=0.50\textwidth}}
  \caption{Completeness for photometrically selecting binary quasars as
    a function of redshift determined by applying the criteria in 
    \S~\ref{sec:selection} $\sim 300,000$ simulated binaries with $i < 21$. 
    \label{fig:bin_complete}}
\end{figure}

The completeness of the binary selection as a function of redshift is
required for the clustering analysis in Paper II. For concreteness,
consider the particular redshift range ${\rm z}=3.9-4.0$, where the
completeness for selecting individual quasars is $77\%$
(Figure~\ref{fig:complete}). Without imposing criterion III), the
completeness for selecting binaries would be $53\%$, the square of the
individual quasar completeness, but with a $4\%$ reduction due to
criterion II). We naively guess that the combined KDE$+$ SDSS $+$FIRST
sample has the same completeness as our photometric selection (for
individual quasars) in this redshift bin. Then requiring that one
member of each pair match these auxiliary catalogs implies that we
will recover $1-(1-0.77)^2 = 94\%$ of the binaries.  Assuming the
objects missed by our photometric selection and the auxiliary catalogs
are uncorrelated, the total resulting completeness for binaries is
$0.94\times 0.53$ or $50\%$. More importantly, we do not need to
characterize the KDE catalog very precisely, since even a $\sim 20\%$
reduction in completeness (highly unlikely) would only amount to
$\lesssim 10\%$ error in characterizing our overall completeness for
binaries. The other sources of statistical and systematic error in the
clustering measurement in Paper II significantly overwhelm this
conservative estimate of the error in our completeness. Thus we make
the Ansatz that the auxiliary catalogs (KDE$+$ SDSS $+$FIRST) have the
same completeness as our individual photometric quasar
selection\footnote{We do not impose criterion J) (see
  \S~\ref{sec:photo}) in this estimate since no photometric redshift
  cuts were made on the KDE catalog.} (see Figure~\ref{fig:complete}).
Figure~\ref{fig:bin_complete} shows the final completeness of our
binary quasar selection as a function of redshift with 
this assumption.

\subsection{Survey Status}
\label{sec:status}

\input table3.tex

\begin{figure}[!t]
  \centerline{\epsfig{file=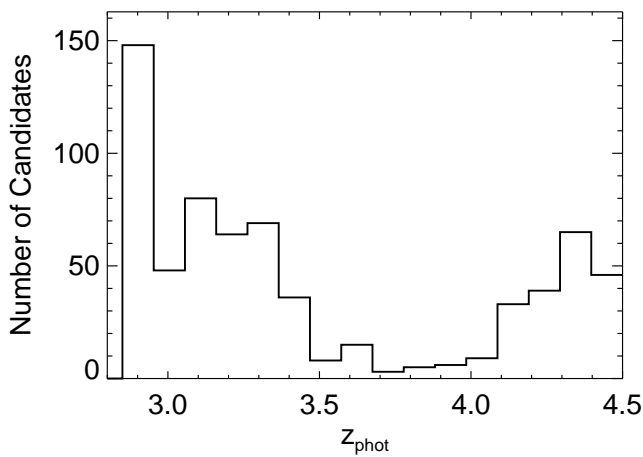,bb=0 0 504 360,
      width=0.50\textwidth}}
  \caption{Distribution of photometric redshifts of both members of the 340 
    quasar pair candidates with $\theta < 60\arcsec$ selected by our survey. 
    \label{fig:status}}
\end{figure}

Our sample of 27 ${\rm z} > 2.9$ binaries (24 of them new discoveries) is
presented in \S~\ref{sec:sample}. Here we discuss the status of
spectroscopic follow-up observations of binary candidates in our
survey.  The redshift and angular distribution of our sample, shown in
Figure~\ref{fig:scatter}, reflects specific biases in our follow-up
observations.  First, we tended to observe small separation pairs
first, because they are more interesting for studies of the
intervening IGM and contribute more significantly to the quasar
clustering signal (see Paper II).  Thus our follow-up is significantly
less complete for larger splittings.  As a result we limit our
clustering analysis in Paper II to $\theta < 60\arcsec$, and the
status of our follow-up observations is quantified only for these
separations. Of the 27 binaries in our sample, 22 have $\theta <
60\arcsec$, but only 15 of these satisfy the selection criteria laid
out in \S~\ref{sec:selection} (the other seven were mostly discovered
as pair candidates from the KDE catalog).  The second noticeable trend
in Figure~\ref{fig:scatter} is that there are very few pairs with
redshifts ${\rm z} = 3.2-3.5$ and at ${\rm z} > 4.1$.  Figure~\ref{fig:status}
shows the distribution of mean photometric redshifts of the 340
(non-spurious) quasar pair candidates with $\theta < 60\arcsec$
selected by our survey. As expected, the largest number of candidates
occur at redshifts where the efficiency is low in
Figure~\ref{fig:eff}, and thus at the corresponding redshifts we find
fewer binaries in Figure~\ref{fig:scatter}, both because fewer
candidates at these redshifts were targeted for follow-up spectroscopy
and also the success rate is much lower.

We ran our binary selection on $8142$~deg$^2$ of SDSS imaging, which
resulted in 431 pair candidates with $\theta < 60\arcsec$. The SDSS
images of all candidate quasar pairs were visually inspected to reject
bad imaging data.  The primary contaminants are regions around
diffraction spikes from bright stars, challenging deblends, and fields
containing a globular cluster where the stellar density is incredibly
high.  Of the 431 initial pair candidates, 91 were spurious, leaving
340 candidate quasar pairs.  For some of the candidates, one member of
the pair was observed by the SDSS spectroscopic survey and the spectra
were examined in these cases to check for mis-identifications; there
were 21 objects which were not ${\rm z} > 2.9$ quasars, leaving 319
candidates for follow-up.  Color-color diagrams for each candidate
were also inspected, and each candidate was assigned a rank (HIGH,
MEDIUM, LOW) for follow-up observations. These rankings are
subjective, reflecting the likelihood that the candidates were quasars
based on their colors, the fidelity of the photometry (noisy data or
suspicious de-blends received a lower priority), and if both members were
members of either the KDE, FIRST, or SDSS spectroscopic catalogs. In addition
smaller separation pairs were given a higher priority.  

The result of a successful spectroscopic follow-up observation of a
quasar pair candidate falls into one of four categories: (1) a
quasar-quasar pair at the same redshift (2) a projected pair of
quasars at different redshifts (3) a quasar-non-quasar pair
(i.e. either a star or a galaxy) (4) a pair of non-quasars.  Of the
319 candidates for spectroscopic follow-up, 47 have one member
confirmed to be a ${\rm z} > 2.9$ quasar by SDSS spectroscopy and 9 have one
member in the FIRST catalog (which was not targeted by the SDSS
spectroscopic survey).  For only 37 of these 319 candidates are both
objects members of the KDE catalog.

The clustering analysis in Paper II splits our binary sample into two
redshift bins ($2.9 < z < 3.5$ and $3.5 < z < 4.5$), and we describe
the status of our follow-up observations similarly in bins of mean
photometric redshift.  Table~\ref{table:status} summarizes the status
of all of the 319 candidates for spectroscopic follow-up.  To date,
our follow-up spectroscopy of binary candidates is about half
finished, with 46 HIGH priority candidates observed and a total of 55
HIGH priority candidates remaining. Including the medium priority
candidates in the tally implies we are only one third finished, but
our experience is that these lower priority candidates have a much
lower success rate.  Of 170 LOW priority targets, 132 have one member
with ${\rm z}_{\rm phot} < 3.2$, indicating high contamination where the
quasar and stellar loci cross at ${\rm z}\sim 2.8$ (see
Figures~\ref{fig:col_plot}, \ref{fig:eff}, and \ref{fig:status}).  We
caution that the fraction of candidates observed thus far is a
poor predictor of the number of new binaries among our remaining
targets, both because we observed the best candidates first and
because the success rate for the lower priority candidates is unknown.


Indeed our tendency to observe the best candidates first is largely
responsible for our high efficiency for finding binary quasars. Out of
the 46 candidates observed to date with $\theta < 60\arcsec$, 16 were
quasars at the same redshift: 15 correspond to binaries published in
Table~\ref{table:sample}\footnote{Note that only 15 of the total 27
  binaries published in Table~\ref{table:sample} satsify both $\theta
  < 60\arcsec$ and the criteria in \S~\ref{sec:photo} and
  \S~\ref{sec:binary}.} and the other was the gravitational lens
SDSS~J1400$+$3134 at ${\rm z}=3.32$ \citep{Inada08}. The remainder of
our follow-up observations resulted in six projected quasar pairs, 15
quasar-non-quasar pairs, and nine pairs of non-quasars.

\section{Observations}

\label{sec:observations}

\input table1.tex

In this section we describe the observations for our high-redshift
binary survey. In \S~\ref{sec:spectro} we discuss the follow-up
spectroscopy of pair candidates and in \S~\ref{sec:zsys} we explain
our algorithm for estimating quasar systemic redshifts. As we
elaborate below, optical spectra covering rest-frame ultraviolet
emission lines yield quasar redshifts with large velocity
uncertainties ($\sim 1000~\kms$). Thus redshifts estimation must be
treated carefully in order to distinguish true binary quasars from
projected pairs.


\subsection{Spectroscopic Observations}
\label{sec:spectro}

Discovery spectra of our binary candidates were obtained with the
Astrophysical Research Consortium 3.5m telescope at the Apache Point
Observatory (APO), the 6.5m Multiple Mirror Telescope (MMT), the
Palomar 200-inch telescope, and the Mayall 4m telescope at Kitt Peak
National Observatory. For the confirmation observations we used
low-resolution gratings and we typically oriented the slit at the
position angle between the two quasars so that both members of the
pair could be observed simultaneously. Exposure times varied with the
telescope and observing conditions, but ranged from 600s for the
bright candidates on the larger telescopes to 2400s for the faintest
on the smaller telescopes. The relevant details of the confirmation
observations are summarized in Table~\ref{table:telescopes}.

A subset of the confirmed binary quasars was later observed at higher
resolution and signal-to-noise ratio using the Echelle Spectrograph
and Imager \citep[ESI,][]{ESI} on the Keck II 10m telescope on Mauna
Kea during a number of runs between November 2006 and July 2008.  The
goal of the ESI observations was to measure small scale transverse
correlations of the Ly$\alpha$ forest and \ion{C}{4} metal line
absorbers to reveal the small scale structure of the IGM
\citep{thesis,ehm+07}, characterize the size of metal-enriched regions
\citep{ehm+07,martin09}, and constrain the dark energy density of the
Universe with the Ly$\alpha$ forest Alcock-Paczy{\'n}ski test
\citep{AP79,MM99,HSB99,Pat03}. Our binary quasar search also uncovered
many high-redshift projected pairs of quasars (see
\S~\ref{sec:status}), which we also observed with ESI to study
foreground quasar environments in absorption \citep{QPQ1,QPQ2,QPQ3}
and the transverse proximity effect
\citep[e.g.][]{Croft04,KT08}.

ESI is an echellette spectrograph with ten orders on the CCD, giving
continuous spectral coverage from 4000\AA-10000\AA~.  The resolving
power is nearly constant across the orders. Our observations
exclusively used the $1.0\arcsec$ slit, resulting in a resolution of
$R\simeq 5000$ or FWHM$\simeq 60\kms$. As the ESI slit is 20$\arcsec$
long, the instrument was rotated to observe both quasars
simultaneously for pairs with $\Delta \theta \lesssim 15\arcsec$,
whereas wider separation pairs were observed individually with the
position angle set to the parallactic angle.  The signal-to-noise
ratio of the resulting spectra vary, but typically peak at ${\rm S\slash N}
\sim 15-20$ per pixel between the Ly$\alpha$ and \ion{C}{4} emission
lines (around 7000\AA) , with lower ${\rm S\slash N}$ at bluer and redder
wavelengths. Total exposure times varied from 1hr for our brighter
targets to 6hr for our faintest, with typical exposures around 2hr.

The low resolution discovery spectra and ESI spectra were reduced
using the
LowRedux\footnote{http://www.ucolick.org/$\sim$xavier/LowRedux/index.html}
and
ESIRedux\footnote{http://www.ucolick.org/$\sim$xavier/ESIRedux/index.html}
data reduction pipelines, respectively, written in the Interactive
Data Language (IDL).

\subsection{Estimating Systemic Redshifts}

\label{sec:zsys}

Quasar redshifts determined from the rest-frame ultraviolet emission
lines (redshifted into the optical at ${\rm z}\gtrsim 3$) can differ by over
one thousand kilometers per second from the systemic frame, because of
outflowing/inflowing material in the broad line regions of quasars
\citep{gaskell82,TF92,vanden01,Richards02,Shen08}. A redshift
determined from the narrow forbidden emission lines, such as
[\ion{O}{3}]~$\lambda 5007$, or Balmer lines such as
H$\alpha$~$\lambda 6563$ are much better predictors of systemic
redshift \citep{boroson05}, but at ${\rm z} \gtrsim 3$, spectral coverage of
these lines would require observations in the near-infrared. Instead,
we follow the approach described in \citet{Shen07}, and determine the
redshifts of quasars from our optical spectra.  \citet{Shen07}
measured the correlation between the relative shifts of the
high-ionization emission lines \ion{Si}{4}, \ion{C}{4},\ion{C}{3}, and
the shift between these respective lines and the systemic frame, as
traced (approximately) by the \ion{Mg}{2} line
\citep{Richards02}. These correlations can then be used to estimate
the systemic redshift and the associated error, from high-ionization
line centers.  As an operational definition, we consider quasar pairs
with velocity differences of $|\Delta v|\leq 2000\kms$ to be at the
same redshift, since this brackets the range of velocity differences
caused by both peculiar velocities, which could be as large as $\sim
500\kms$ if binary quasars reside in rich environments, and systemic
redshift uncertainties, which can be larger than $\sim 1000~\kms$.



Our quasar spectra come from a variety of instruments (see
Table~\ref{table:telescopes}) and have varying signal-to-noise ratios
and spectral coverage. The ESI spectra are always used for redshift
determination when available, otherwise we use SDSS spectra or the
low signal-to-noise ratio discovery spectra. For noisy spectra which
can have intrinsic absorption features, flux weighted centering often
yields erroneous results for emission line centers. For this reason,
we adopt the more robust line-centering procedure introduced in
\citet{QPQ1}, and used by \citet{Shen07}. For those spectra where
prominent UV emission lines cannot be centered because of low
signal-to-noise ratio, limited spectral coverage, or strong intrinsic
absorption features, the redshift is measured using the peak of the
Ly$\alpha$ emission line measured after smoothing with a $1000\kms$
boxcar. A shift is then applied between the Ly$\alpha$ redshift and
the systemic frame, which was measured on average by \citet{Shen07}.

\section{The Binary Quasar Sample}
\label{sec:sample}
\input table2.tex


\begin{figure}[!t]
  \centerline{\epsfig{file=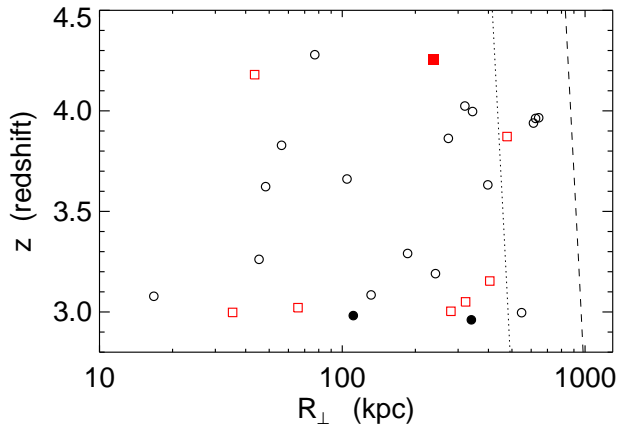,bb=70 70 576 432,width=0.50\textwidth}}
  \caption{Distribution of redshift and proper transverse separation
    for 27 binary quasars in the SDSS imaging footprint. The (black)
    circles indicate the subset of objects which meet the selection
    criteria in \S~\ref{sec:selection}, whereas the (red) squares do
    not. Previously known binaries are indicated by filled
    symbols. The dotted and dashed curves are the proper transverse
    distance corresponding to the fiber collision limit of the SDSS
    ($\theta=55\arcsec$), and the angular separation limit of our
    binary search ($\theta=120\arcsec$),
    respectively. \label{fig:scatter}}
\end{figure}

We present a sample of 24 new high redshift, $2.9 \lesssim z \lesssim
4.3$, binary quasars with (proper) transverse separations $ 10~{\rm
  kpc} < R_{\perp} < 650~{\rm kpc}$. Eight members of this sample are
very close binaries with $R_{\perp} < 100~{\rm kpc}$, and four of these are at
${\rm z} > 3.5$. Table~\ref{table:sample} lists relevant quantities for a
total of 27 binaries in the SDSS imaging area footprint, including three
previously known pairs. These are SDSSJ~1116$+$4118 and
SDSSJ~1546$+$5134, which were published in \citet{BINARY}, and
SDSSJ~1439$-$0033, discovered serendipitously by \citet{Schneider00}.
Of this sample of binaries, 22 of 27 have angular separations $\theta
\le 55\arcsec$, below the SDSS fiber collision scale.  For two of the
binaries both objects were observed by the SDSS spectroscopic survey:
SDSSJ~1016$+$4040 has angular separation $\theta = 68.2\arcsec$,
i.e. larger than the fiber collision limit, and SDSSJ~1116$+$4118
($\theta = 13.8\arcsec$) was observed by overlapping spectroscopic
plates \citep{BINARY}.

Several of the binary quasars listed in Table~\ref{table:sample} do
not meet all of the criteria laid out in \S~\ref{sec:selection}. As we
describe further below, PSS~1315$+$2924, one of the closest systems in
our sample, was selected by a different method altogether as its faint
companion is not even detected in the SDSS imaging data. The remaining
binaries which fail to meet these criteria are: SDSSJ~0004$-$0844,
SDSSJ~0829$+$2927, SDSSJ~1150$+$4659, SDSSJ~1159$+$3426,
SDSSJ~1248$+$1957, and SDSSJ~1307$+$0422, as well as SDSSJ~1439$-$0033
discovered by \citet{Schneider00}. Many of these objects were targeted
for follow-up because they were pair candidates in the KDE photometric
catalog. In Paper II, we restrict the
clustering analysis to the subset of pairs which meet the selection
criteria in \S~\ref{sec:selection}, for which we calibrated the
completeness of our search.

The distribution of redshifts and proper transverse separations probed
by the 27 binary quasars in the SDSS footprint is illustrated by the
scatter plot in Figure~\ref{fig:scatter}. The relative paucity of
pairs with $\theta \gtrsim 60\arcsec$ results from the incompleteness
of our spectroscopic follow-up campaign.  The lack of pairs with ${\rm
  z} = 3.2-3.5$ and at ${\rm z} > 4.1$ occurs because photometric
quasar selection is inefficient at these redshifts (see discussion in
\S~\ref{sec:status}).

Spectra of selected members of our binary sample are shown in
Figures~\ref{fig:spec3} ($2.9 < z < 3.5$) and \ref{fig:spec4} (${\rm z} >
3.5$). All of the spectra in these figures are from ESI with the
exception of SDSSJ~1116$+$4118, for which we show the SDSS
spectra. 

\begin{figure*}
  \centerline{\epsfig{file=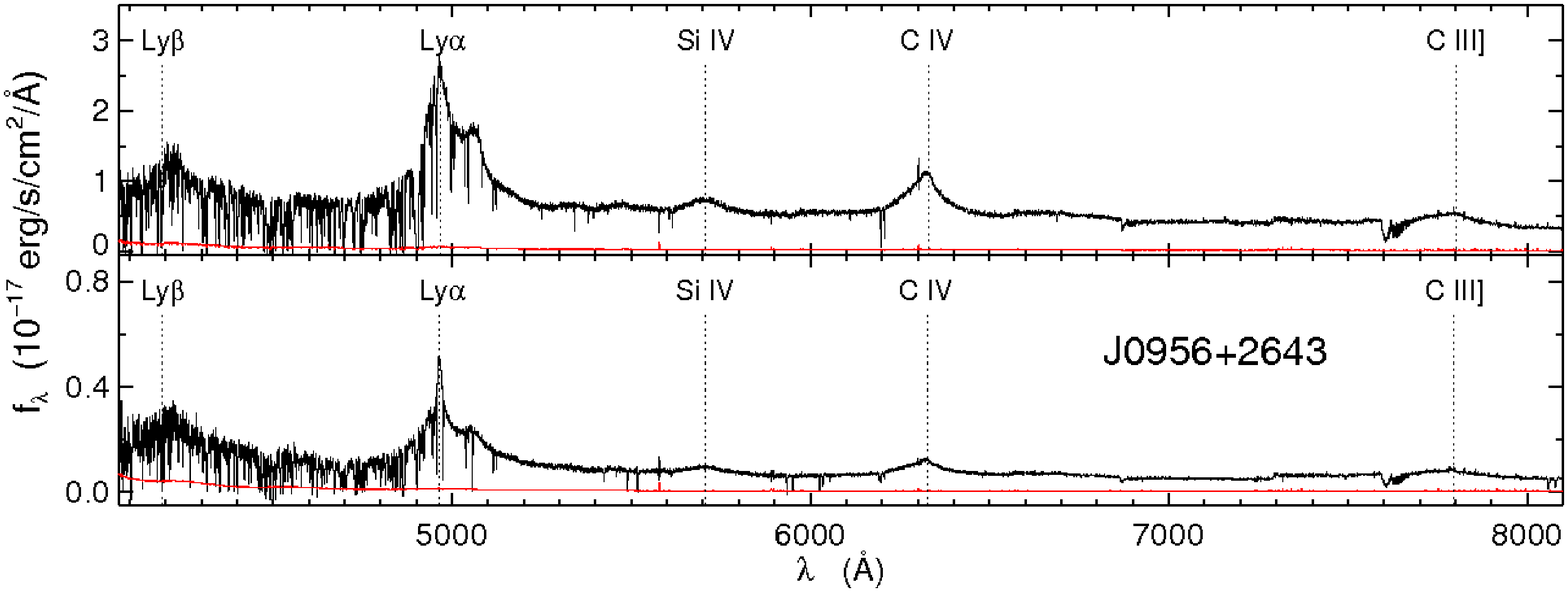,bb=0 -10 936 360,width=\textwidth}}
  \centerline{\epsfig{file=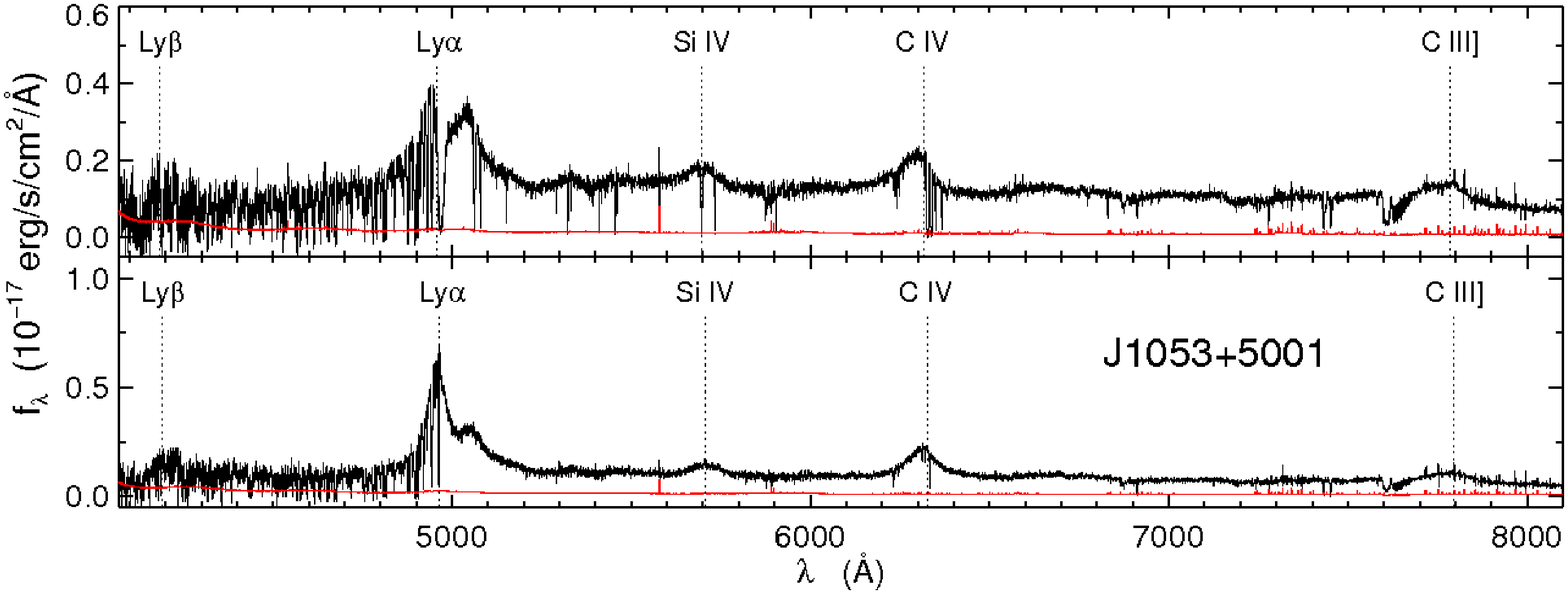,bb=0 -10 936 360,width=\textwidth}}
  \centerline{\epsfig{file=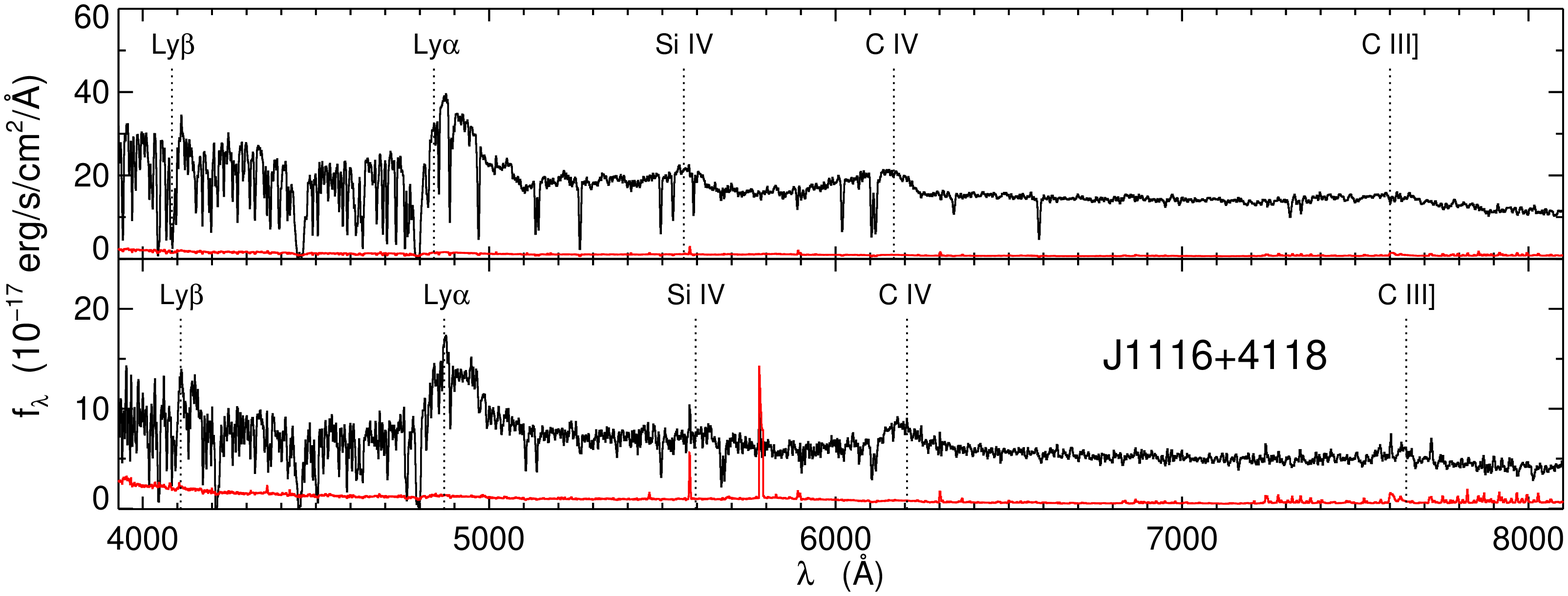,bb=0 -10 936 360,width=\textwidth}}
  \caption{ Spectra of selected binary quasars in the redshift range
    $2.9 < z < 3.5$. A three pixel boxcar smoothing has been applied
    to all spectra. The red curves indicate the 1$-\sigma$ error array. 
    Object `A' is always shown in the top panel of each plot. 
    All spectra are from ESI with the exception of SDSSJ~1116$+$4118,
    for which we show the SDSS spectra. The strong features in the ESI
    spectra at $7600$\AA~are atmospheric telluric absorption.}
\end{figure*}
\addtocounter{figure}{-1}
\begin{figure*}
  \centerline{\epsfig{file=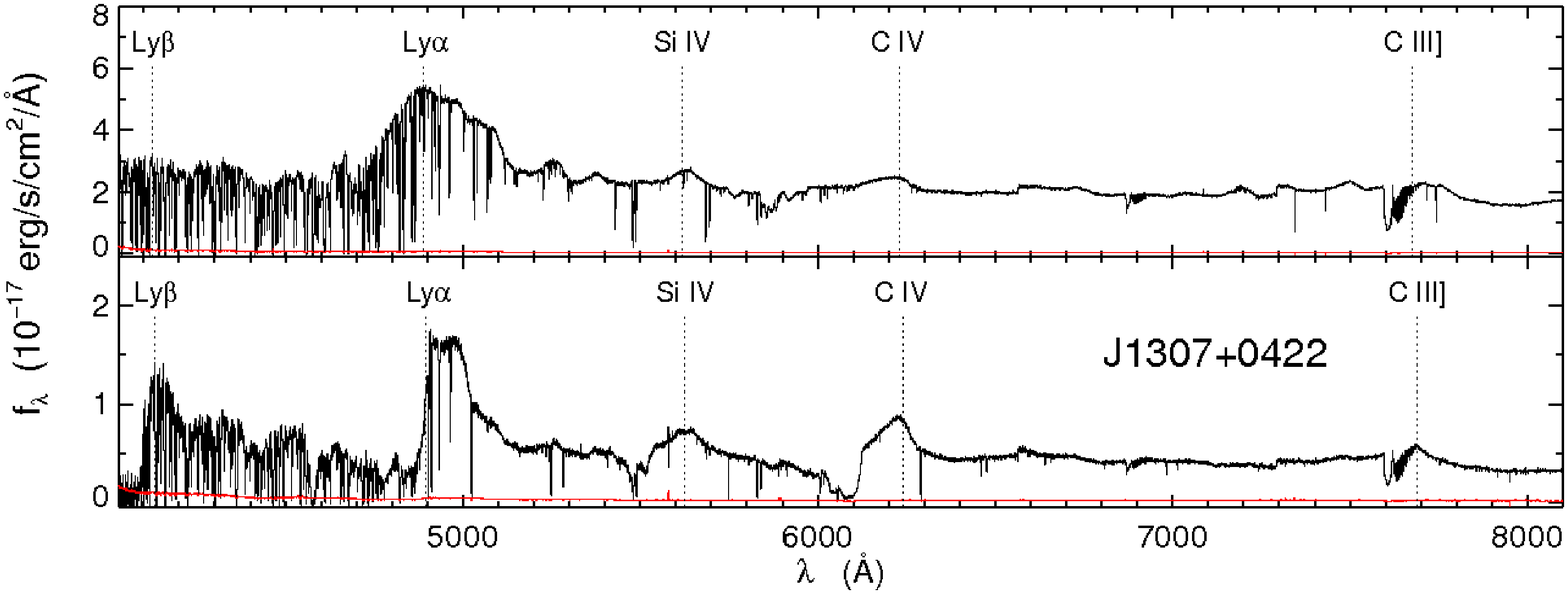,bb=0 -10 936 360,width=\textwidth}}
  \centerline{\epsfig{file=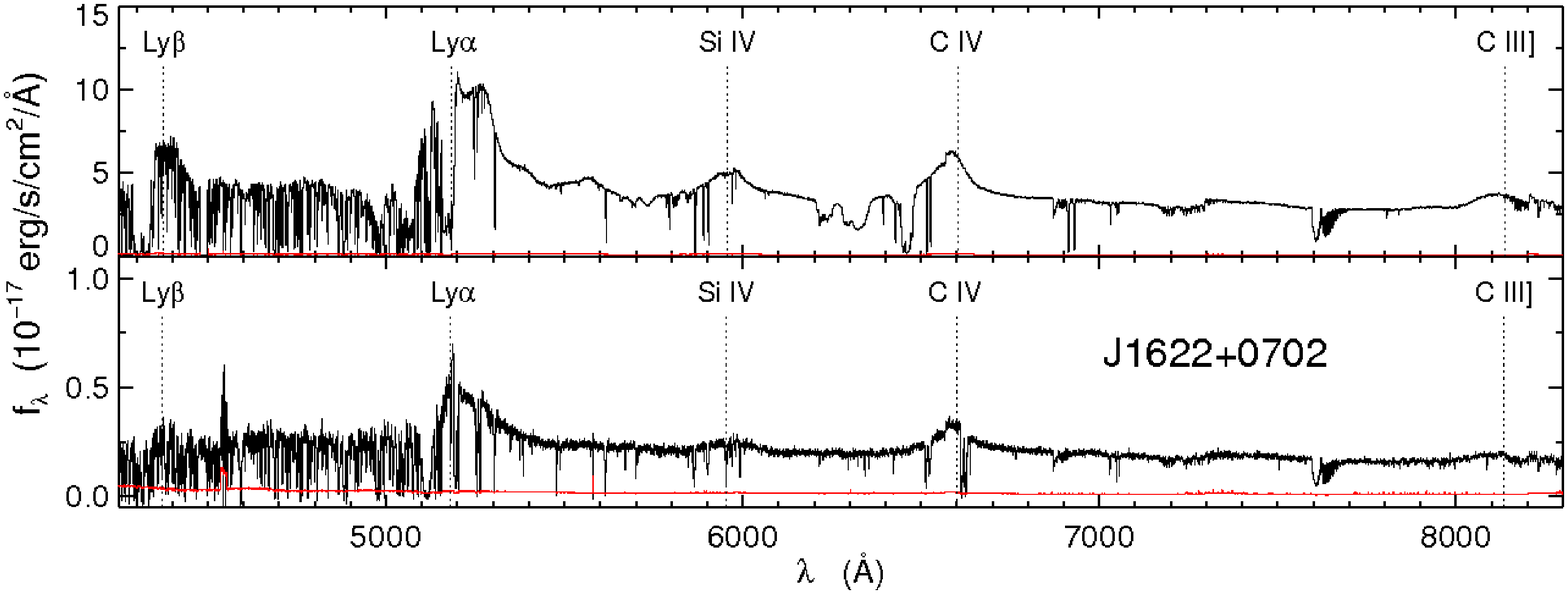,bb=0 -10 936 360,width=\textwidth}}
  \caption{continued.}\label{fig:spec3}
\end{figure*}


\begin{figure*}
  \centerline{\epsfig{file=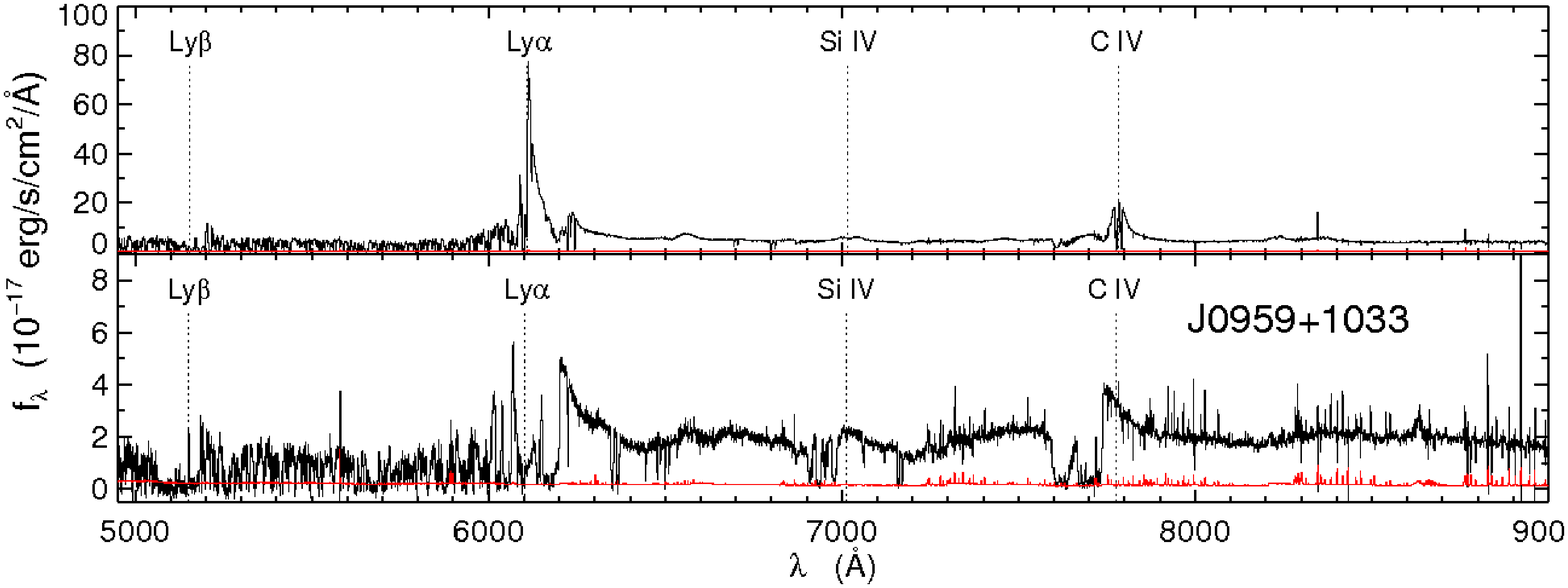,bb=0 -10 936 360,width=\textwidth}}
  \centerline{\epsfig{file=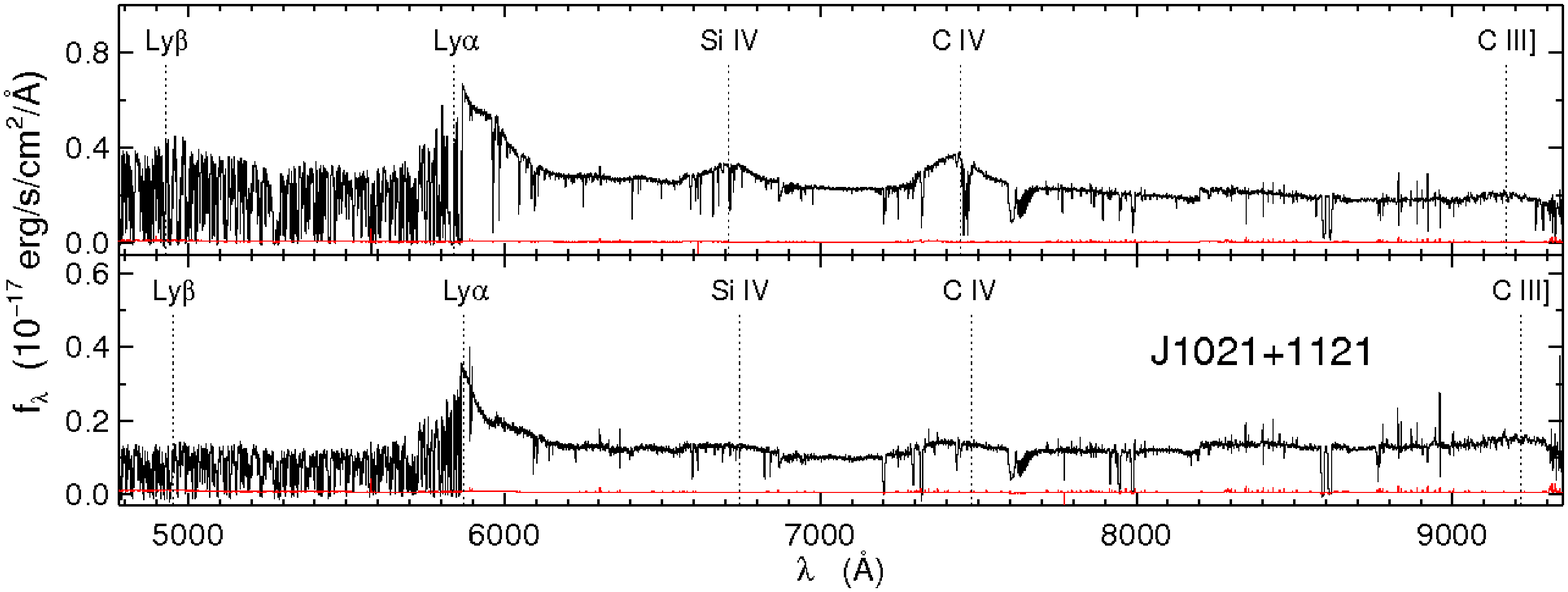,bb=0 -10 936 360,width=\textwidth}}
  \centerline{\epsfig{file=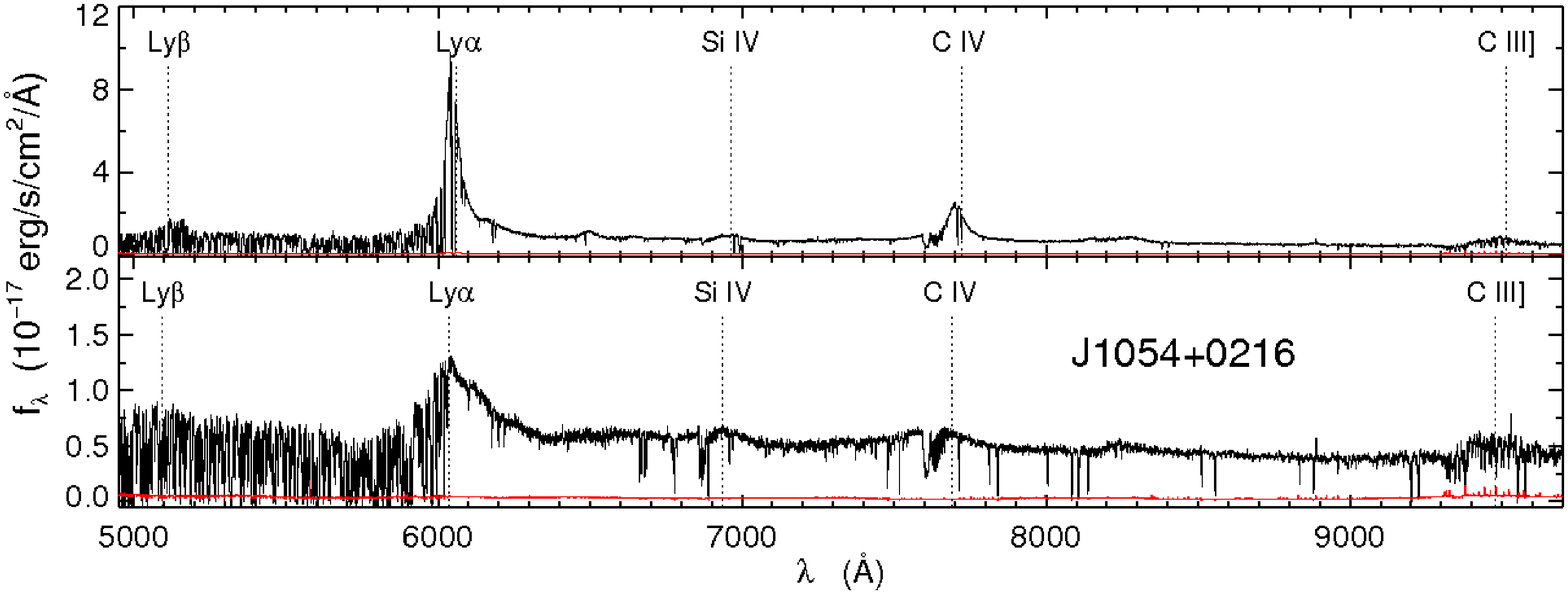,bb=0 -10 936 360,width=\textwidth}}
  \caption{Same as Figure~\ref{fig:spec3} but for binary quasars with
    ${\rm z} > 3.5$.}
\end{figure*}
\addtocounter{figure}{-1}
\begin{figure*}
  \centerline{\epsfig{file=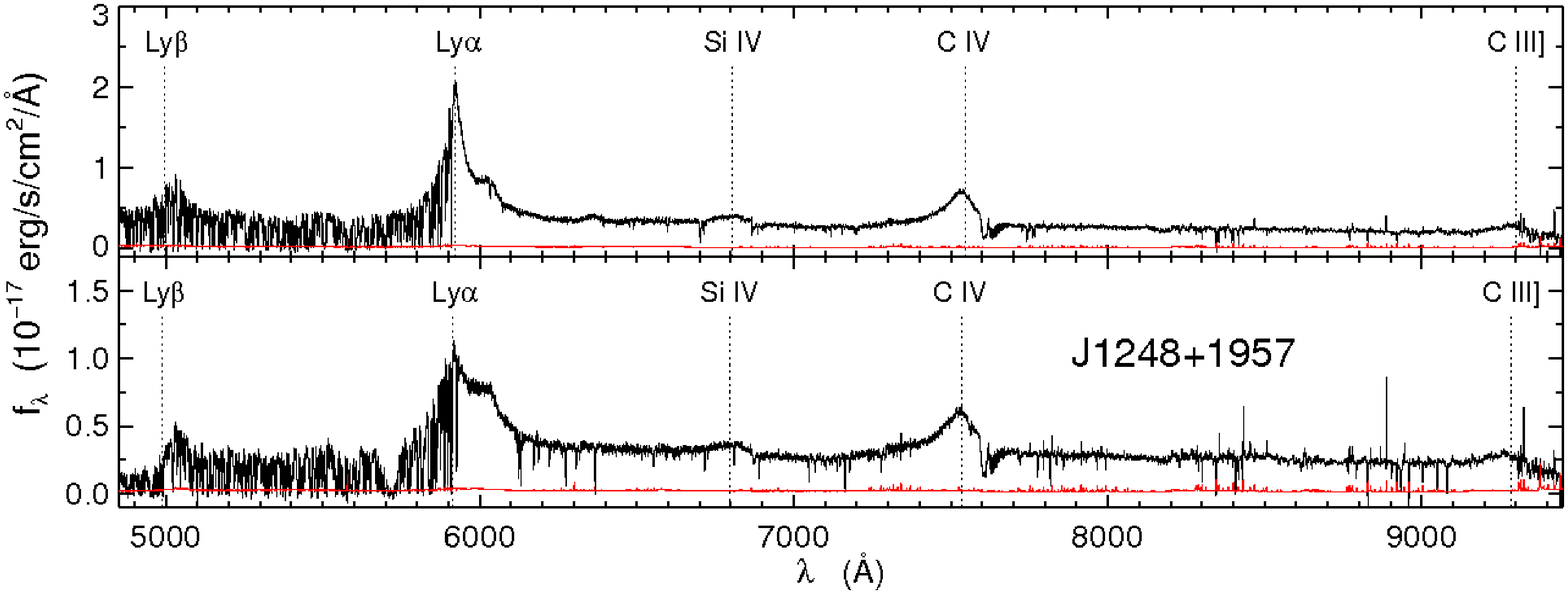,bb=0 -10 936 360,width=\textwidth}}
  \centerline{\epsfig{file=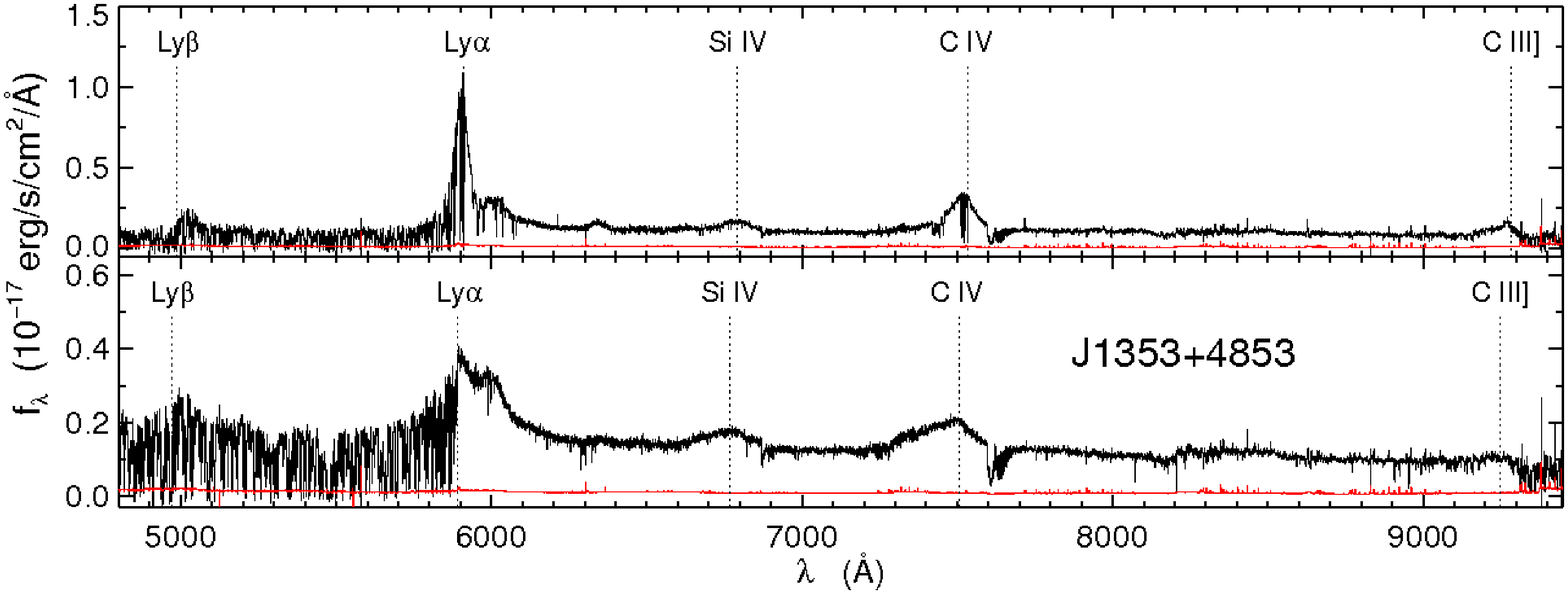,bb=0 -10 936 360,width=\textwidth}}
  \centerline{\epsfig{file=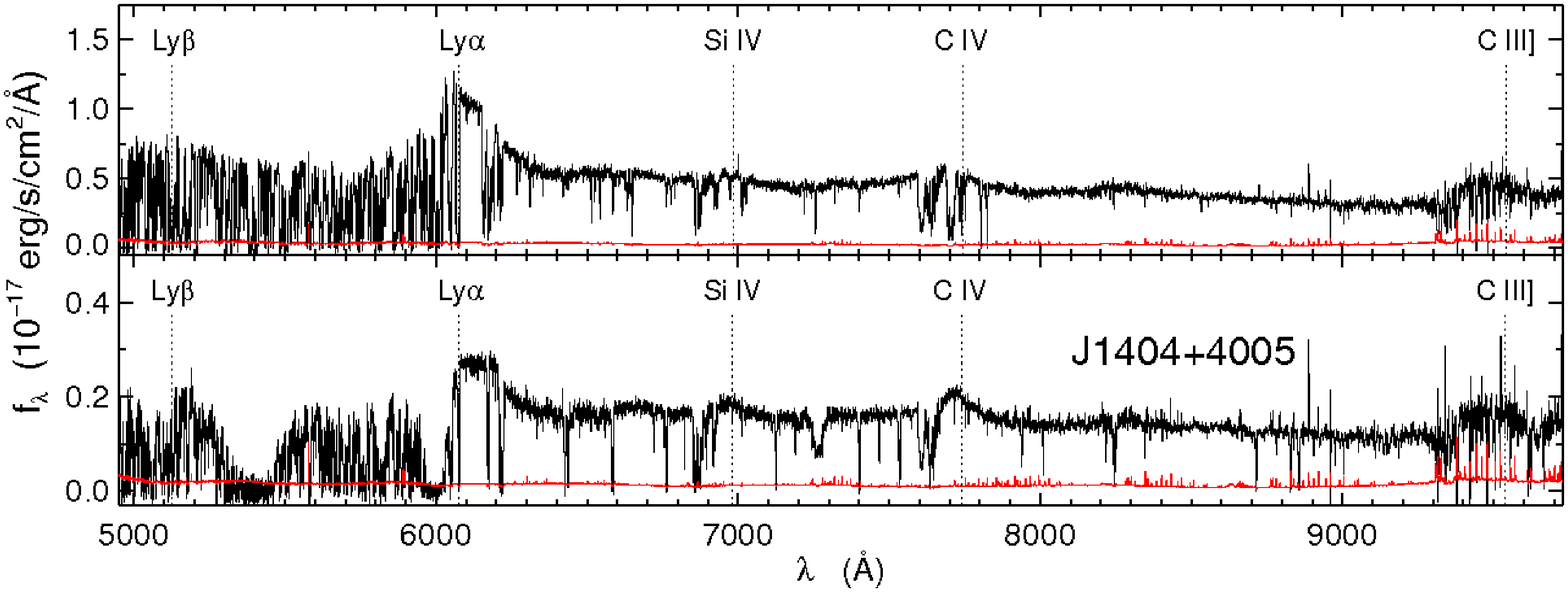,bb=0 -10 936 360,width=\textwidth}}
  \caption{continued.}
\end{figure*}
\addtocounter{figure}{-1}
\begin{figure*}
  \centerline{\epsfig{file=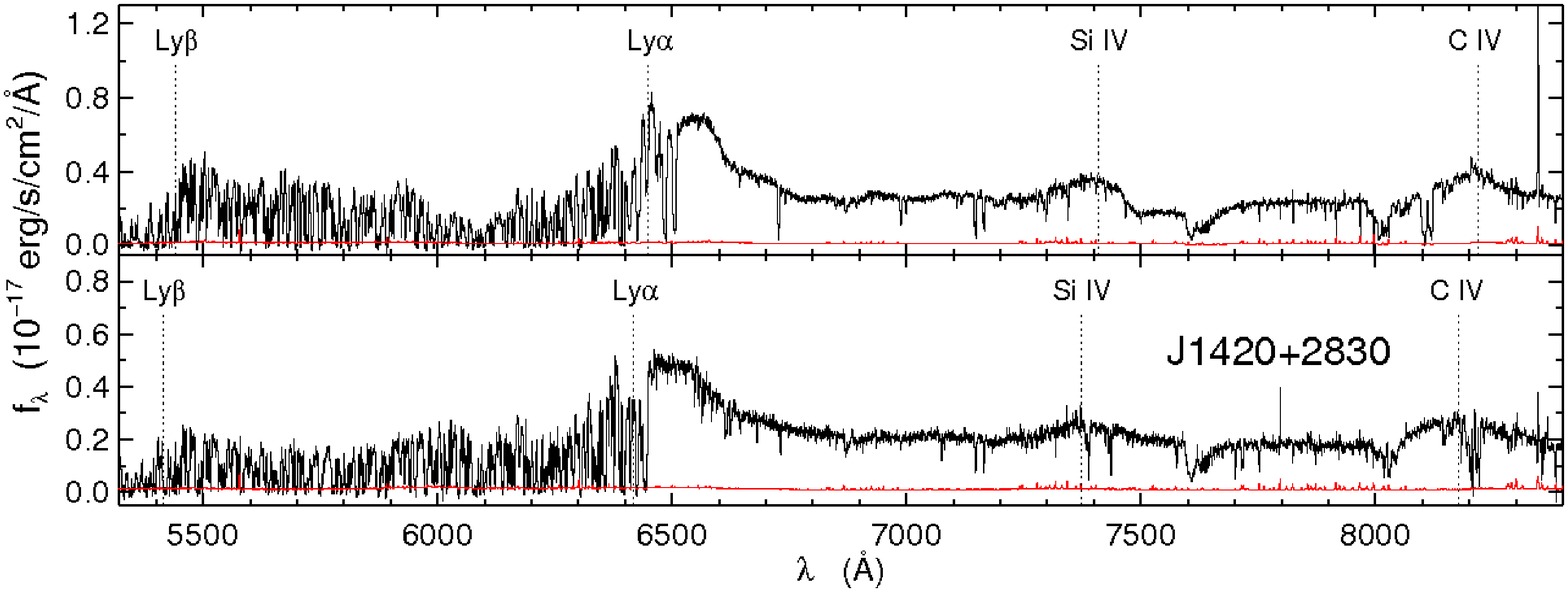,bb=0 -10 936 360,width=\textwidth}}
  \centerline{\epsfig{file=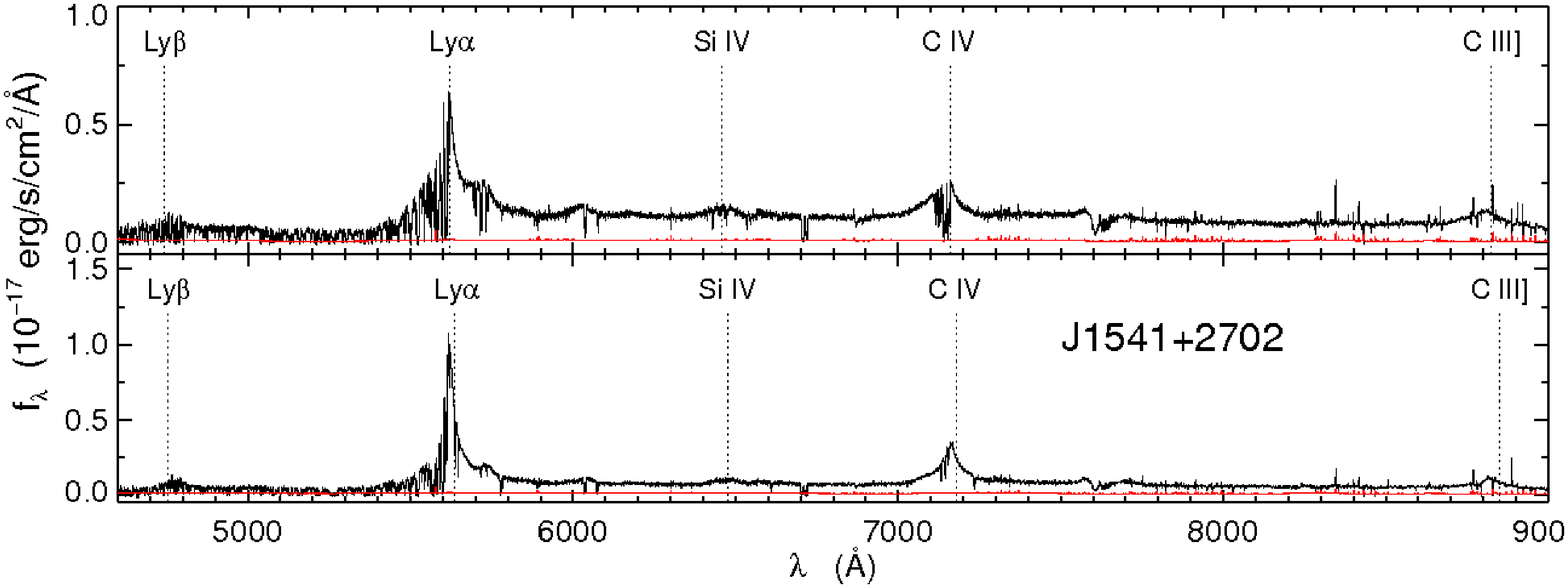,bb=0 -10 936 360,width=\textwidth}}
  \caption{continued.}\label{fig:spec4}
\end{figure*}

\subsection{The Discovery of PSS~1315+2924}
\label{sec:pss}

The faint quasar companion to the bright quasar PSS~1315$+$2924 was
discovered during a pilot search for proto-clusters associated with ${\rm z}
> 4$ quasars \citep{Djor99a,Djor99b}. Two different strategies were
employed in this search. In the first, $R$-band snapshot images were
taken of about twenty ${\rm z} \sim 4$ quasars to moderate depth $R
\lesssim 25$ using the Low Resolution Imaging Spectrograph
\citep[LRIS;][]{LRIS} on the Keck I telescope. If a companion was found
within $\sim 10\arcsec$ of the quasar it was spectroscopically
observed with LRIS in longslit mode. For the second approach, deep
imaging was obtained in BRI ($R \lesssim 26$) for a sample of about 
10 ${\rm z}\sim 4$ quasars, again using LRIS.  Color-selection was used to
search for dropout candidates (at these redshifts, the continuum drop
is dominated by the Ly$\alpha$ forest, rather than the Lyman break,
which is used to select galaxies at ${\rm z}\sim 2-3.5$). These candidates
were then followed up with multislit spectroscopy using LRIS in
spectroscopic mode. The faint $R \simeq 24.1$ companion
PSS~1315$+$2924B (see Table~\ref{table:sample}) was found from this
color-selection approach, and spectra of both quasars were obtained on
14 April 1999 UT with LRIS, and reduced using standard techniques. In
addition, this proto-cluster survey uncovered about two dozen
star-forming galaxies at small angular separations $\theta \lesssim
10\arcsec$ from the quasars. A similar survey was also carried out
around ${\rm z}\sim 5-6$ quasars, which led to the discovery of a companion
${\rm z}=5.02$ quasar $196\arcsec$ away from a high redshift quasar at
${\rm z}=4.96$ \citep{Djor03}.

\subsection{Contamination by Gravitational Lenses}

It is possible that some of the quasar pairs with image splittings
$\lesssim 20^{\prime\prime}$ in our sample are wide separation strong
gravitational lenses rather than binary quasars. Indeed, several
gravitationally lensed quasars with image splittings $\theta \gtrsim
3\arcsec$ have been discovered in the SDSS
\citep{Quad03,Quad04,Oguri05,Inada08}. In particular, our binary
search recovered one of the high-redshift gravitational lenses from
the sample of \citet{Inada08}, SDSS~J1400$+$3134 at ${\rm z}=3.32$
with image splitting $\theta = 1.7\arcsec$. The largest separation
lens discovered to date is the ${\rm z}=2.2$ quasar lensed into two
images separated by $22.5\arcsec$ by a foreground cluster
\citep{Inada06}. But the lower number density of ${\rm z} > 3$ quasars
make high redshift wide-separation lenses correspondingly rarer than
at ${\rm z}\sim 2$. Using cosmological ray-tracing simulations,
\citet{QSOLENS} determined that a quasar sample with the flux limit of
our search should contain $\sim 2$ lenses with image splitting $\theta
> 10\arcsec$ and ${\rm z} > 3$.  However, this estimate is extremely
sensitive to the assumed value of $\sigma_8$ \citep{Li06,Li07}, which
was high ($\sigma_8=0.95$) for the \citet{QSOLENS} study; the
expectation for high-redshift lenses is therefore lower by a factor of
$\gtrsim 5$ \citep{Li07} if one adopts the lower value of $\sigma_8 =
0.80$ favored by the WMAP five-year data \citep{wmap05}.

A quasar pair can be positively confirmed as a binary if the spectra
of the images are vastly different \citep[c.f. ][]{Gregg02}, if only
one of the images is radio-loud \citep[an $O^2R$ pair, in the notation
of ][]{Koch99}, or if the quasars' hosts are detected and they are not
clearly lensed. The sufficient conditions for a pair to be identified
as a lens are the presence of more than two images in a lensing
configuration, the measurement of a time delay between images, the
detection of a plausible deflector, or the detection of lensed host
galaxy emission \citep[see e.g.][]{Koch99,MWF99}. 

The sample of binaries in Table~\ref{table:sample} contains ten
systems with $\theta < 20\arcsec$ which we consider as potential
lenses. We have ESI spectra of all but two of these,
SDSSJ~0004$-$0844\footnote{A spectrum of SDSSJ~0004$-$0844 was
  obtained the Magellan Echellette (MagE) Spectrograph on the Magellan
  Clay Telescope. These data are not published here, but the different
  Ly$\alpha$ emission line profiles also allow us to compellingly
  argue against gravitational lensing.} and PSSJ~1315$+$2924. For all
cases with ESI spectra, we can convincingly rule out the lensing
hypothesis because of their spectral dissimilarity (see
Figures~\ref{fig:spec3} and \ref{fig:spec4}). In addition,
SDSSJ~1307$+$0422A was detected in the FIRST survey with 20cm flux
$F_{\rm A}=14.3~{\rm mJy}$. Using $i$-band apparent magnitudes to
compute a flux ratio implies the flux of the second image would be
$F_{\rm B}=3.8~{\rm mJy}$ under the lensing hypothesis. This is four
times brighter than the FIRST flux limit, however, no radio
counterpart is detected\footnote{Although the optical flux ratio could
  be anomalous relative to the radio because of microlensing,
  anomalies as large as a factor of $\sim 4$ are rare
  \citep[e.g.][]{Pooley07}.}.

For both SDSSJ~0004$-$0844 (${\rm z}=3.00$, $\theta = 4.4\arcsec$) and
PSSJ~1315$+$2924 (${\rm z}=4.18$, $\theta = 6.1\arcsec$) we have obtained
deep imaging using the Low Resolution Imaging Spectrograph
\citep[LRIS;][]{LRIS} on the Keck I telescope. For SDSSJ~0004$-$0844
our LRIS $R$-band image reaches an approximate point source depth of
$R \sim 24$, and there is no evidence for a lensing galaxy or cluster.
Our LRIS image of PSSJ~1315$+$2924 reaches a depth of $R\sim 26$, and
there is similarly no evidence for a lens.  Although the
\emph{absence} of a deflector in images of a quasar pair does not
strictly speaking confirm the binary hypothesis, it makes it very
likely. For wide separation lenses with $\Delta\theta\gtrsim3\arcsec$,
such as SDSSJ1004+4112 ($\Delta\theta_{\rm max}=14.62\arcsec$, ${\rm z}_{\rm
  lens}=0.68$), Q~0957+561 \citep[$\Delta\theta=6.2\arcsec$, ${\rm z}_{\rm
  lens}=0.36$;][]{Walsh79}, or SDSSJ~1029$+$2623 ($\Delta\theta_{\rm
  max}=22.5\arcsec$, ${\rm z}_{\rm lens}\sim 0.6$) where the lens is a
bright galaxy in a cluster or group, the lens galaxies would have $r
\sim 20-21$ at the most probable redshifts of ${\rm z}=0.3-0.7$
\citep{QSOLENS}, and would thus be easily detectable in our LRIS
images. Although we cannot completely rule out lenses with ${\rm z}\gtrsim
1$, a wide separation ${\rm z} > 3$ quasar lensed by such a high redshift
group or cluster is extremely unlikely because structures with
appreciable strong-lensing cross-section are predicted to be extremely
rare at ${\rm z} > 1$ \citep{QSOLENS,Hennawi07}.  Thus we are confident that
all of the quasar pairs in our sample are indeed binary quasars.

\section{Summary and Conclusions}
\label{sec:conc}

We have conducted the first systematic survey to discover high
redshift binary quasars. Using color-selection and photometric
redshift techniques, we mined $8142$~deg$^2$ of SDSS imaging data for
binary quasar candidates, and confirmed them spectroscopically with
follow-up observations on 4m class telescopes.  Our sample of 24 new
binaries at redshifts $2.9 \lesssim z \lesssim 4.3$ with (proper)
transverse separations $ 10~{\rm kpc} < R_{\perp} < 650~{\rm kpc}$
increases the number of such objects known by an order of magnitude.
Eight members of this sample are very close pairs with $R_{\perp} < 100~{\rm
  kpc}$, and of these close systems four are at ${\rm z} > 3.5$. More
importantly, whereas the only previously known close binary at
${\rm z}\sim 4$ was discovered serendipitously \citep{Schneider00}, we
defined an algorithm for homogeneously selecting binary candidates and
quantified its completeness and efficiency using simulations. 

Armed with this new sample of binaries and a well-understood selection
function, Paper II presents the first measurement of the small-scale
clustering ($R < 1~\hMpc$ comoving) of high-redshift quasars. On
large scales, high-redshift quasars cluster much more strongly than
their low-redshift counterparts \citep{Shen07}, indicating that
quasars are hosted by extremely massive dark matter halos ($M\gtrsim
10^{13}~\hmsol$) --- perhaps not so surprising as these quasars likely
harbor the most extreme black holes ($M\gtrsim 10^{9}~\msol$) in the
Universe \citep[e.g.][]{Shen08}. Small-scale clustering measurements
characterize the $10~{\rm kpc}-1~{\rm Mpc}$ scale environments of
these extreme black holes, providing important clues about how quasar
activity is triggered \citep{Thacker06,Hopkins08,Wetzel09a,Thacker08}
and constraining how quasars populate their dark matter halos
\citep{Wetzel09a,Shankar09}.  Because our survey is only about half
complete, we can only place lower limits on the small-scale
clustering. Nevertheless, even these lower limits provide important
constraints on models (see Paper II).


The massive dark halos inhabited by high-redshift quasars
\citep{Shen08} suggests that high-redshift binaries represent
exponentially rare conjunction of two extremely massive black holes
(and possibly two dark matter halos).  Furthermore, the simultaneity
of their luminous quasar phases is another important
coincidence. Using the luminosity function and the number of binaries
detected in our survey, we can estimate the fraction of quasars which
have an active companion. In the redshift range $3.5 < z < 4.5$, there
are $N_{\rm QSO}(i < 21) = 1.9~{\rm deg^{-2}}$, or about $\sim 15,000$
in the SDSS area we surveyed. We uncovered four binary quasars with $R
< 100~{\rm kpc}$ ($\theta \lesssim 15\arcsec$) and $i < 21$, which
given our incompleteness (see Figure~\ref{fig:bin_complete}), implies
$\sim 8$ binaries actually exist. Thus the companion probability is
extremely small $\sim 0.05\%$. Equivalently, over this redshift range
our survey volume is $\sim 100~{\rm Gpc}^3$ (comoving), implying a
comoving number density $n_{\rm binary} \sim 10^{-10}~{\rm Mpc}^{-3}$, or
about one binary per $10~{\rm Gpc}^3$. For comparison, this is an
order of magnitude rarer than the extremely rare ${\rm z} \sim 6$
quasars \citep{Fan01}, and corresponds to a volume thirty times larger
than the Millennium Simulation \citep{Springel05}. Despite their
paucity, the handful of binaries presented here amount to a
substantial small-scale clustering signal \citep{Shen09}: although the
observed pair fraction is incredibly low, the probability of such
associations occurring at random is orders of magnitude lower.



Binaries are scarce in our survey partly because at ${\rm z} = 4$, $i < 21$
corresponds to an extremely luminous (and therefore rare) quasar, $M_i
< -26.5$\footnote{Following \citet{Richards06} we quote $i$-band
  absolute magnitudes K-corrected to ${\rm z}=2$.}, i.e. comparable to
3C~273 and about $\sim 3.5$ magnitudes brighter than the canonical
distinction ($M_{\rm B} < -23$) between quasars and Seyferts. Upcoming
deep synoptic surveys such as the Panoramic Survey Telescope \& Rapid
Response System \citep[Pan-STARRS][]{PanSTARRS} and the Large Synoptic
Survey Telescope \citep[LSST][]{LSST}, will combine variability and
color-selection to identify large photometric samples of faint quasars
with high completeness and efficiency. The number of binaries
uncovered by these surveys scales as $N \propto N_{\rm QSO}^2A$, where
$N_{\rm QSO}$ is the quasar number counts and $A$ is the survey area.
Extrapolating our luminosity function to $i<24.5$ (i.e. $M_i < -23$ at
${\rm z} = 4$), we find $N_{\rm QSO}(i<24.5) \simeq 30~{\rm deg^{-2}}$ for
$3.5 < z < 4.5$, or about 15 times the number density of the SDSS in
this range. The LSST will discover all active galaxies over
20,000~${\rm deg^{-2}}$ even beyond this depth, implying a factor of
$\sim 600$ increase in the number of quasar pairs. If we naively
assume clustering is independent of luminosity \footnote{The
  luminosity dependence of large-scale quasar clustering at
  high-redshift has yet to be quantified, although there is tentative
  evidence that the brightest quasars cluster more strongly at ${\rm z}
  \lesssim 2$ \citep{Shen09}.} and extrapolate from the size of our
sample, we predict $\sim 5000$ binary quasars with $R \lesssim
100~{\rm kpc}$ and $3.5 < z < 4.5$ in the LSST imaging. The shallower
($i < 24$) but wider ($A = 3\pi$) Pan-STARRS1 survey will contain a
comparable number. The clustering measurement errors scale as the
square-root of the number of binaries, implying errors about 25 times
smaller than achieved in \citep{Shen09} from our sample. Furthermore,
the fainter sample and dramatically improved statistics will allow a
precise investigation of the redshift and luminosity dependence of
both small and large scale clustering, resulting in powerful
constraints on clustering models.


Searching for galaxy overdensities around the binaries published here
will shed light on their environments and the mechanism triggering
their quasar activity.  Although the separations of our binary quasars
are too large for direct gravitational interactions to trigger
accretion, they could reside in the same overdense environments whereby
interactions with other galaxies make quasar activity more likely
\citep[e.g.][]{Djor91,Hopkins08}.  The idea of using quasars as
beacons to pinpoint dense regions in the early Universe is not new
\citep[e.g.][]{Djor99a,Djor99b}. Deep imaging studies of high-redshift
quasar environments in search of overdensities and proto-clusters have
been conducted at ${\rm z}\sim 3$ \citep[e.g.][]{Steidel99,Infante03}, by
Djorgovski and collaborators at ${\rm z} \sim 4$
\citep{Djor99a,Djor99b,Djor03}, and by several groups at ${\rm z}\sim 5-6$,
\citep{Stiavelli05,Zheng06,Kashikawa07,Kim08,Priddey08}. While it is
still unclear whether high-redshift quasars reside in proto-clusters,
compelling evidence for overdensities has been found around a handful
of ${\rm z} > 4$ radio galaxies
\citep[e.g][]{Venemans02,Venemans04,Miley04,Overzier06,Venemans07,Overzier08}.
If similar deep imaging and multi-object spectroscopy of our high
redshift binaries uncovers proto-cluster environments, this would
establish that either the binaries are the highest sigma peaks in the
early Universe, or that enhanced quasar activity is triggered by
galaxy interactions. Furthermore, the kinematics and size of such an
overdensity could distinguish these alternatives.  Intriguingly,
\citet{FS04} obtained deep Subaru images of the ${\rm z}=4.25$ quasar
pair discovered by \citet{Schneider00} and found no evidence for an
overdensity of galaxies. But the tenfold increase in high-redshift
binaries published here provides ample opportunity for future detailed
statistical studies.

We conclude with the reminder that our survey for quasar pairs is
ongoing and less than $50\%$ complete.  Although the binary yield will
likely drop because we observed the best candidates first, we
anticipate discovering $10-20$ more high-redshift binaries in the
SDSS imaging.

\acknowledgments 

We acknowledge helpful discussions with J. Cohn, N. Padmanabhan,
D. Schlegel, D. Weinberg, A. Wetzel, and M. White.  For part of this
work JFH was supported by a NASA Hubble Fellowship grant \# 01172.01-A
and by the NSF Postdoctoral Fellowship program (AST-0702879). YS and
MS acknowledge support from the National Science Foundation (NSF)
(AST-0707266). DPS similarly acknowledges support from the NSF
(AST-0607634). SGD, AAM, and EG acknowledge partial support from the
NSF grant AST-0407448 and from the Ajax Foundation. CLM acknowledges
support from the Packard Foundation and the NSF (AST-0808161).

Funding for the SDSS and SDSS-II has been provided by the Alfred
P. Sloan Foundation, the Participating Institutions, the National
Science Foundation, the U.S. Department of Energy, the National
Aeronautics and Space Administration, the Japanese Monbukagakusho, the
Max Planck Society, and the Higher Education Funding Council for
England. The SDSS Web Site is http://www.sdss.org/.

The SDSS is managed by the Astrophysical Research Consortium for the
Participating Institutions. The Participating Institutions are the
American Museum of Natural History, Astrophysical Institute Potsdam,
University of Basel, University of Cambridge, Case Western Reserve
University, University of Chicago, Drexel University, Fermilab, the
Institute for Advanced Study, the Japan Participation Group, Johns
Hopkins University, the Joint Institute for Nuclear Astrophysics, the
Kavli Institute for Particle Astrophysics and Cosmology, the Korean
Scientist Group, the Chinese Academy of Sciences (LAMOST), Los Alamos
National Laboratory, the Max-Planck-Institute for Astronomy (MPIA),
the Max-Planck-Institute for Astrophysics (MPA), New Mexico State
University, Ohio State University, University of Pittsburgh,
University of Portsmouth, Princeton University, the United States
Naval Observatory, and the University of Washington.

Some of the data presented herein were obtained at the W.M. Keck
Observatory, which is operated as a scientific partnership among the
California Institute of Technology, the University of California and
the National Aeronautics and Space Administration. The Observatory was
made possible by the generous financial support of the W.M. Keck
Foundation.  The authors wish to recognize and acknowledge the very
significant cultural role and reverence that the summit of Mauna Kea
has always had within the indigenous Hawaiian community.  We are most
fortunate to have the opportunity to conduct observations from this
mountain.



\bibliographystyle{apj}
\bibliography{allrefs}

\end{document}

%% file: defs.tex
\newcommand{\beq}{\begin{equation}}
\newcommand{\eeq}{\end{equation}}

\def\be{\begin{equation}}
\def\ee{\end{equation}}
\def\bea{\begin{eqnarray}}
\def\eea{\end{eqnarray}}



\def \logTd6 {\hbox{log$( T/6 \kev)$} }

\def\myputfigure#1#2#3#4#5%
{\vskip#5pt\makebox[0pt]{\hskip#2in
\includegraphics[width=#3\textwidth]{#1}}\vskip#4pt\hfill}

\def \arcsec    {^{\prime\prime}}

\def \kms            {~{\rm km~s}^{-1}}

\def \etal      {et al.}

\def \kev       {{\rm\ keV}}
\def \msol      {{\rm\ M}_\odot}
\def \hmsol     {h^{-1}{\rm\ M}_\odot}
\def \hMpc      {h^{-1}{\rm\ Mpc}}
\def \hkpc      {h^{-1}{\rm\ kpc}}

%% file: table3.tex
\begin{deluxetable}{l|c|c|c}
\tablecolumns{3}
\tablewidth{0pc}
\tablecaption{Status of Quasar Pair Candidates with $\Delta\theta < 60\arcsec$
\label{table:status}}
\tablehead{Status & ${\bar z_{\rm phot}} < 3.5$ & ${\bar z_{\rm phot}} > 3.5$ & Total} 
\startdata
Observed & 17  & 29  & 46\\
HIGH     & 28  & 27  & 55\\
MEDIUM   & 16  & 32  & 48\\
LOW      & 156 & 14  & 170\\
\tableline 
Total    & 217 & 102 & \phn319
\enddata
\tablecomments{The number of quasar pair candidates are listed and
  photometric redshifts of both members were averaged, allowing us to
  split the candidates into two redshift bins about ${\bar z_{\rm
      phot}}=3.5$. Binary candidates which were targeted for follow-up
  observations are labeled as 'Observed'. The remaining candidates
  have a priority of HIGH, MEDIUM, or LOW.}
\end{deluxetable}

%% file: table1.tex
\begin{deluxetable*}{lcclclc}
\tablecolumns{15}
\tablewidth{0pc}
\tablecaption{Binary Quasar Spectroscopy\label{table:telescopes}}
\tablehead{Telescope  & Instrument    & Spectrograph Type & Spectral Coverage & FWHM    & Dates & Reference}
\startdata
              APO 3.5m  & DIS             &       Double           & \ \ \  $3800-10,000$ & 440/270 & Nov. - Dec., 2005 & 1 \\
                MMT     & Red Channel     &       Single           & \ \ \ $5000-10,000$  &  1000   & Feb. 19-21, 2006  & 2 \\ 
                MMT     & Blue Channel    &       Single           & \ \ \ $3200-8400$   &   600 & Dec. 13-15, 2006 and Mar. 25, 2007 & 3 \\ 
       Palomar 200-inch & DBSP            &       Double           & \ \ \ $3100-9300$  & 900/550 & Sep. 2007 - Apr. 2009 & 4\\ 
              Mayall 4m & RC-Spectrograph &       Single           & \ \ \ $3600-9200$  &   325 & Feb. 9-11, 2008 and Jun. 7-10, 2008 & 5\\
            Keck II 10m & ESI             &       Echellette       & \ \ \ $4000-10,000$  &    60   & Nov. 2006 - Jul. 2008 & 6\\
\enddata
\tablecomments{(1) ? (2) ? (3) ? (4) \citet{DBSP} (5) ? (6) \citet{ESI}}
\end{deluxetable*}

%% file: table2.tex
\begin{deluxetable*}{rcccccccccccccc}
\tablecolumns{15}
\tablewidth{0pc}
\tablecaption{High Redshift Binary Quasars in the SDSS Imaging Footprint
\label{table:sample}}
\tablehead{Name & RA & Dec & $u$ & $g$ & $r$ & $i$ & $z$ & $\chi_{\rm star}^2$ & $\chi_{\rm phot}^2$ & z & $\sigma_{\rm z}$ & $\Delta v$ & $\Delta \theta$ & $R_{\perp}$ }
\startdata
\hfill SDSS\,J0004$-$0844A &   00 04 50.674 &   $-$08 44 49.48 &   21.89  &   20.75  &   20.56  &   20.57  &   20.67  &   \phn\phn1.2  &   \phn\phn0.9  &   2.998  &   \phn1450  &   \phn\phn650  &   \phn4.4  &   \phn35 \\
\smallskip
\hfill SDSS\,J0004$-$0844B &   00 04 50.920 &   $-$08 44 51.97 &   22.80  &   21.02  &   20.87  &   20.84  &   20.66  &   \phn\phn4.6  &   \phn\phn0.3  &   2.989  &   \phn1450             &             &             & \\
\hfill SDSS\,J0829$+$2927A &   08 29 07.722 &   $+$29 27 54.71 &   23.64  &   20.93  &   20.28  &   19.68  &   19.30  &   \phn63.3  &   \phn22.5  &   3.054  &   \phn1450  &   \phn\phn270  &    40.3  &    322 \\
\smallskip
\hfill SDSS\,J0829$+$2927B &   08 29 09.018 &   $+$29 27 18.10 &   22.04  &   20.07  &   19.95  &   19.83  &   19.88  &   \phn36.8  &   \phn\phn1.2  &   3.050  &   \phn\phn710             &             &             & \\
\hfill SDSS\,J0956$+$2643A &   09 56 27.149 &   $+$26 43 24.47 &   21.61  &   19.43  &   19.22  &   19.28  &   19.30  &    143.2  &   \phn\phn2.6  &   3.087  &   \phn\phn520  &   \phn\phn150  &    16.5  &    131 \\
\smallskip
\hfill SDSS\,J0956$+$2643B &   09 56 25.934 &   $+$26 43 21.62 &   22.71  &   20.78  &   20.46  &   20.46  &   20.32  &   \phn14.1  &   \phn\phn0.6  &   3.085  &   \phn\phn520             &             &             & \\
\hfill SDSS\,J0959$+$1032A &   09 59 03.843 &   $+$10 32 45.26 &   23.16  &   20.92  &   19.23  &   19.12  &   19.19  &    215.2  &   \phn\phn3.5  &   4.024  &   \phn1450  &   \phn\phn250  &    44.1  &    320 \\
\smallskip
\hfill SDSS\,J0959$+$1033B &   09 59 05.139 &   $+$10 33 25.02 &   25.86  &   21.47  &   20.00  &   19.56  &   19.50  &   \phn24.5  &   \phn\phn1.7  &   4.020  &   \phn1500             &             &             & \\
\hfill SDSS\,J1016$+$4040A &   10 16 01.509 &   $+$40 40 52.89 &   21.26  &   19.42  &   19.22  &   19.10  &   18.84  &    109.8  &   \phn\phn8.1  &   2.996  &   \phn\phn520  &   \phn1030  &    68.2  &    548 \\
\smallskip
\hfill SDSS\,J1016$+$4040B &   10 16 05.842 &   $+$40 40 05.80 &   22.12  &   20.36  &   20.06  &   20.10  &   19.92  &   \phn49.7  &   \phn\phn1.8  &   2.983  &   \phn1450             &             &             & \\
\hfill SDSS\,J1021$+$1112A &   10 21 16.982 &   $+$11 12 27.55 &   25.23  &   21.99  &   20.50  &   20.35  &   20.18  &   \phn26.7  &   \phn\phn0.3  &   3.805  &   \phn\phn520  &   \phn1470  &   \phn7.6  &   \phn56 \\
\smallskip
\hfill SDSS\,J1021$+$1112B &   10 21 16.468 &   $+$11 12 27.83 &   25.46  &   22.36  &   20.62  &   20.73  &   20.26  &   \phn58.2  &   \phn\phn5.4  &   3.829  &   \phn\phn630             &             &             & \\
\hfill SDSS\,J1053$+$5001A &   10 53 20.150 &   $+$50 01 46.02 &   23.77  &   21.05  &   20.72  &   20.72  &   20.64  &   \phn18.2  &   \phn\phn0.4  &   3.078  &   \phn\phn630  &   \phn\phn490  &   \phn2.1  &   \phn17 \\
\smallskip
\hfill SDSS\,J1053$+$5001B &   10 53 20.041 &   $+$50 01 47.84 &   23.35  &   21.10  &   20.84  &   20.88  &   21.02  &   \phn19.4  &   \phn\phn1.5  &   3.085  &   \phn\phn520             &             &             & \\
\hfill SDSS\,J1054$+$0215A &   10 54 34.174 &   $+$02 15 51.95 &   23.03  &   20.22  &   18.69  &   18.61  &   18.75  &    244.9  &   \phn\phn6.2  &   3.984  &   \phn\phn630  &   \phn1120  &    88.3  &    644 \\
\smallskip
\hfill SDSS\,J1054$+$0216B &   10 54 28.515 &   $+$02 16 16.37 &   23.21  &   20.58  &   19.41  &   19.10  &   18.97  &   \phn12.9  &   \phn\phn5.3  &   3.966  &   \phn\phn790             &             &             & \\
\hfill SDSS\,J1116$+$4118A &   11 16 11.740 &   $+$41 18 21.51 &   20.30  &   18.49  &   18.13  &   17.93  &   17.95  &   \phn87.0  &   \phn\phn6.0  &   2.982  &   \phn\phn520  &   \phn1840  &    13.8  &    111 \\
\smallskip
\hfill SDSS\,J1116$+$4118B &   11 16 10.689 &   $+$41 18 14.44 &   21.29  &   19.41  &   19.14  &   19.01  &   19.02  &   \phn87.9  &   \phn\phn1.7  &   3.006  &   \phn\phn520             &             &             & \\
\hfill SDSS\,J1118$+$0202A &   11 18 22.764 &   $+$02 02 36.35 &   24.82  &   21.13  &   19.81  &   19.77  &   19.86  &   \phn75.0  &   \phn\phn0.8  &   3.939  &   \phn\phn790  &   \phn\phn\phn\phn0  &    83.8  &    613 \\
\smallskip
\hfill SDSS\,J1118$+$0201B &   11 18 17.920 &   $+$02 01 54.48 &   26.78  &   21.83  &   20.62  &   20.56  &   20.22  &   \phn27.3  &   \phn\phn8.3  &   3.939  &   \phn2000             &             &             & \\
\hfill SDSS\,J1150$+$4659A &   11 50 55.751 &   $+$46 59 41.96 &   22.43  &   20.52  &   20.34  &   20.56  &   20.16  &   \phn66.7  &   \phn12.2  &   3.005  &   \phn\phn710  &   \phn\phn140  &    34.9  &    280 \\
\smallskip
\hfill SDSS\,J1150$+$4659B &   11 50 52.353 &   $+$46 59 39.44 &   23.24  &   21.17  &   20.79  &   20.91  &   20.91  &   \phn26.7  &   \phn\phn3.4  &   3.003  &   \phn\phn790             &             &             & \\
\hfill SDSS\,J1159$+$3426A &   11 59 06.632 &   $+$34 26 48.24 &   23.16  &   20.34  &   20.00  &   19.93  &   19.87  &   \phn39.0  &   \phn\phn0.0  &   3.135  &   \phn\phn790  &   \phn1340  &    51.2  &    405 \\
\smallskip
\hfill SDSS\,J1159$+$3427B &   11 59 02.974 &   $+$34 27 12.31 &   24.35  &   21.21  &   20.97  &   21.15  &   21.07  &   \phn21.9  &   \phn\phn5.7  &   3.154  &   \phn\phn790             &             &             & \\
\hfill SDSS\,J1248$+$1957A &   12 48 46.012 &   $+$19 57 16.86 &   22.55  &   20.41  &   19.48  &   19.44  &   19.61  &   \phn81.7  &   \phn12.1  &   3.872  &   \phn\phn520  &   \phn\phn480  &    64.8  &    477 \\
\smallskip
\hfill SDSS\,J1248$+$1956B &   12 48 42.237 &   $+$19 56 39.90 &   24.39  &   21.75  &   20.16  &   20.01  &   20.10  &   \phn91.8  &   \phn\phn1.2  &   3.864  &   \phn\phn520             &             &             & \\
\hfill SDSS\,J1251$+$2715A &   12 51 23.788 &   $+$27 15 24.61 &   23.14  &   21.29  &   20.29  &   20.21  &   20.02  &   \phn19.7  &   \phn\phn5.0  &   3.660  &   \phn1500  &   \phn\phn\phn90  &    13.9  &    105 \\
\smallskip
\hfill SDSS\,J1251$+$2715B &   12 51 22.746 &   $+$27 15 23.60 &   23.22  &   21.82  &   20.78  &   20.68  &   20.57  &   \phn14.6  &   \phn\phn3.5  &   3.661  &   \phn1450             &             &             & \\
\hfill SDSS\,J1307$+$0422A &   13 07 56.734 &   $+$04 22 15.57 &   20.22  &   18.18  &   17.97  &   17.59  &   17.43  &    140.4  &   \phn24.0  &   3.021  &   \phn\phn710  &   \phn\phn490  &   \phn8.2  &   \phn66 \\
\smallskip
\hfill SDSS\,J1307$+$0422B &   13 07 56.184 &   $+$04 22 15.50 &   21.41  &   19.68  &   19.25  &   19.02  &   19.12  &   \phn18.3  &   \phn\phn9.1  &   3.028  &   \phn\phn710             &             &             & \\
\hfill SDSS\,J1312$+$4616A &   13 12 40.873 &   $+$46 16 47.86 &   25.60  &   21.78  &   20.18  &   19.86  &   19.46  &   \phn49.7  &   \phn\phn8.4  &   3.971  &   \phn1450  &   \phn\phn530  &    85.7  &    625 \\
\smallskip
\hfill SDSS\,J1312$+$4615B &   13 12 45.032 &   $+$46 15 33.85 &   25.03  &   22.17  &   20.43  &   20.25  &   20.31  &   \phn99.8  &   \phn\phn1.7  &   3.963  &   \phn2000             &             &             & \\
\hfill PSSJ1315$+$2924A$^\dagger$ &   13 15 39.566 &   $+$29 24 40.76 &   23.00  &   21.53  &   19.71  &   19.33  &   19.35  &    113.4  &   \phn\phn3.4  &   4.18  &   \phn2000  &   \phn\phn\phn--  &   \phn6.1  &   \phn44\\
\smallskip
\hfill PSSJ1315$+$2924B$^\dagger$ &   13 15 39.122 &   $+$29 24 38.96 &    \phn--    &   \phn--    &  $\sim$24 &    \phn--    &     \phn--   &    \phn--    &      \phn\phn--        &   4.18  &   \phn2000  &             &             &  \\
\hfill SDSS\,J1353$+$4852A &   13 53 28.570 &   $+$48 52 34.06 &   24.65  &   21.71  &   20.40  &   20.40  &   20.18  &   \phn62.7  &   \phn\phn1.9  &   3.863  &   \phn\phn520  &   \phn1130  &    37.1  &    273 \\
\smallskip
\hfill SDSS\,J1353$+$4853B &   13 53 29.953 &   $+$48 53 08.53 &   25.23  &   21.93  &   20.88  &   20.86  &   20.81  &   \phn23.7  &   \phn\phn1.4  &   3.845  &   \phn\phn520             &             &             & \\
\hfill SDSS\,J1404$+$4005A &   14 04 23.046 &   $+$40 05 11.03 &   24.07  &   20.90  &   19.57  &   19.29  &   19.22  &   \phn68.5  &   \phn\phn1.6  &   3.999  &   \phn1450  &   \phn\phn140  &    47.3  &    344 \\
\smallskip
\hfill SDSS\,J1404$+$4005B &   14 04 19.647 &   $+$40 05 37.79 &   24.98  &   22.47  &   20.97  &   20.50  &   20.47  &   \phn18.8  &   \phn\phn1.1  &   3.997  &   \phn\phn630             &             &             & \\
\hfill SDSS\,J1414$+$3955A &   14 14 04.169 &   $+$39 55 43.54 &   20.73  &   18.72  &   18.57  &   18.64  &   18.58  &    264.8  &   \phn\phn3.1  &   3.218  &   \phn\phn520  &   \phn1950  &    30.7  &    242 \\
\smallskip
\hfill SDSS\,J1414$+$3955B &   14 14 06.766 &   $+$39 55 36.27 &   24.39  &   21.31  &   20.90  &   20.86  &   20.93  &   \phn24.1  &   \phn\phn1.3  &   3.191  &   \phn\phn520             &             &             & \\
\hfill SDSS\,J1420$+$2831A &   14 20 23.773 &   $+$28 31 06.57 &   25.43  &   22.18  &   20.30  &   19.93  &   19.63  &   \phn60.8  &   \phn\phn3.3  &   4.305  &   \phn\phn710  &   \phn1470  &    10.9  &   \phn77 \\
\smallskip
\hfill SDSS\,J1420$+$2830B &   14 20 23.799 &   $+$28 30 55.72 &   23.13  &   23.13  &   20.73  &   20.22  &   20.08  &   \phn65.2  &   \phn\phn5.1  &   4.279  &   \phn\phn790             &             &             & \\
\hfill SDSS\,J1439$-$0033A &   14 39 52.579 &   $-$00 33 59.08 &   25.14  &   23.11  &   20.79  &   20.44  &   20.22  &   \phn53.6  &   \phn\phn3.2  &   4.255  &   \phn\phn\phn\phn0  &   \phn\phn170  &    33.4  &    237 \\
\smallskip
\hfill SDSS\,J1439$-$0034B &   14 39 51.600 &   $-$00 34 29.12 &   23.11  &   23.36  &   21.76  &   21.32  &   20.99  &   \phn\phn4.8  &   \phn\phn3.6  &   4.258  &   \phn\phn\phn\phn0             &             &             & \\
\hfill SDSS\,J1541$+$2702A &   15 41 10.367 &   $+$27 02 24.86 &   24.66  &   21.39  &   20.48  &   20.41  &   20.58  &   \phn35.9  &   \phn\phn2.7  &   3.623  &   \phn\phn630  &   \phn\phn810  &   \phn6.4  &   \phn48 \\
\smallskip
\hfill SDSS\,J1541$+$2702B &   15 41 10.403 &   $+$27 02 31.22 &   23.47  &   21.51  &   20.42  &   20.42  &   20.44  &   \phn52.6  &   \phn\phn4.6  &   3.636  &   \phn\phn520             &             &             & \\
\hfill SDSS\,J1546$+$5134A &   15 46 14.229 &   $+$51 34 05.00 &   20.68  &   19.39  &   19.09  &   18.93  &   18.81  &   \phn11.5  &   \phn\phn5.4  &   2.961  &   \phn1450  &   \phn1240  &    42.2  &    340 \\
\smallskip
\hfill SDSS\,J1546$+$5134B &   15 46 10.538 &   $+$51 34 29.44 &   22.17  &   20.49  &   20.31  &   20.25  &   20.63  &   \phn31.4  &   \phn\phn4.8  &   2.944  &   \phn\phn650             &             &             & \\
\hfill SDSS\,J1622$+$0702A &   16 22 10.109 &   $+$07 02 15.34 &   21.76  &   17.43  &   17.03  &   16.85  &   16.82  &    849.0  &   \phn\phn1.7  &   3.264  &   \phn\phn790  &   \phn\phn160  &   \phn5.8  &   \phn45 \\
\smallskip
\hfill SDSS\,J1622$+$0702B &   16 22 09.815 &   $+$07 02 11.53 &   25.60  &   20.68  &   20.16  &   20.03  &   20.02  &   \phn31.7  &   \phn\phn1.7  &   3.262  &   \phn\phn630             &             &             & \\
\hfill SDSS\,J1626$+$0904A &   16 26 42.067 &   $+$09 04 37.35 &   25.42  &   21.46  &   20.50  &   20.49  &   20.37  &   \phn23.7  &   \phn\phn2.4  &   3.632  &   \phn1450  &   \phn\phn450  &    52.7  &    398 \\
\smallskip
\hfill SDSS\,J1626$+$0904B &   16 26 38.623 &   $+$09 04 24.20 &   25.03  &   21.67  &   20.74  &   20.68  &   20.95  &   \phn18.6  &   \phn\phn2.8  &   3.639  &   \phn1450             &             &             & \\
\hfill SDSS\,J1630$+$1152A &   16 30 55.956 &   $+$11 52 29.43 &   21.80  &   19.33  &   18.93  &   18.71  &   18.66  &   \phn71.2  &   \phn\phn4.4  &   3.277  &   \phn1450  &   \phn\phn970  &    23.8  &    186 \\
\smallskip
\hfill SDSS\,J1630$+$1152B &   16 30 56.727 &   $+$11 52 50.34 &   26.36  &   20.63  &   20.31  &   20.35  &   20.12  &   \phn81.7  &   \phn\phn8.0  &   3.291  &   \phn\phn790             &             &             & \\
\enddata
\tablecomments{Units of right ascension are hours, minutes, and
  seconds, and units of declination are degrees, arcminutes, and
  arcseconds. The brightest member of the pair in $i$-band is always
  labeled `A' and the fainter member is labeled
  `B'. Extinction-corrected SDSS five-band PSF photometry is given in
  the columns $u$, $g$, $r$, $i$, and $z$.  The minimum stellar locus
  and quasar locus distances are denoted by the columns labeled
  $\chi_{\rm star}^2$ and $\chi_{\rm phot}^2$, respectively. The
  redshifts are indicated by the column labeled ${\rm z}$,
  $\sigma_{\rm z}$ is our estimate of the redshift uncertaintiy in
  $\kms$, and $\Delta v$ is the velocity difference between the two
  members of a pair in $\kms$. The column labeled $\Delta \theta$ is
  the angular separation in arcseconds and $R_{\perp}$ is the
  transverse proper separation in kpc. Note that $z$ denotes the
  $z$-band magnitude, whereas z indicates redshift.\\
  $^\dagger$ This quasar pair was selected via a different method
  during a pilot search for proto-clusters associated with $z > 4$
  quasars (see \S~\ref{sec:pss}). The redshifts quoted are approximate
  and we did not measure a velocity difference. The fainter quasar
  PSS~1315$+$2924B is not detected in the SDSS imaging, its quoted
  $r$-band magnitude quoted is an estimate from Keck LRIS $R$-band
  imaging.
}

\end{deluxetable*}